\documentclass[11pt,a4paper]{article}
\usepackage{amsmath,amssymb,amsthm,amsfonts}

\newtheorem{remark}{Remark}
\usepackage{mathrsfs}
\usepackage{xcolor-material}
\usepackage{graphicx}
\usepackage[colorlinks=true,allcolors=blue]{hyperref}
\usepackage[margin=1in]{geometry}
\usepackage{booktabs}
\usepackage[linesnumbered,ruled,vlined]{algorithm2e}
\usepackage{bbm}
\usepackage{pifont}
\newcommand\ph{$\phantom{1}$}
\usepackage{multirow}
\usepackage{enumitem}
\usepackage{xurl}
\usepackage{stackengine}
\usepackage[title]{appendix}
\usepackage{authblk}
\usepackage[round,sectionbib]{natbib}

\usepackage{bibunits}
\defaultbibliographystyle{asa}
\defaultbibliography{ref}

\usepackage{geometry}
\newtheorem{thm}{Theorem}

\newtheorem{prop}{Proposition}

\theoremstyle{definition}

\newtheorem{exmp}{Example}
\newtheorem{asmp}{Assumption}

\theoremstyle{remark}

\def\cX{\mathcal{X}}
\def\cY{\mathcal{Y}}
\def\E{\mathrm{E}}
\def\T{\top}
\def\P{\mathrm{Pr}}
\def\bR{\mathbb{R}}
\def\cov{\mathrm{Cov}}

\def\1{I}
\def\cN{\mathcal{N}}

\def\cls{\mathrm{Sup}}
\def\hat{\widehat}
\def\tilde{\widetilde}
\def\tY{\tilde{Y}}

\def\h{\mathfrak{h}}
\def\cL{\mathcal{L}}

\def\cI{\mathcal{I}}
\def\cJ{\mathcal{J}}
\def\cP{\mathcal{P}}

\def\EL{{\mathrm{EPI}}}
\def\EPI{{\text{EPI }}}

\title{\bf\Large Empirical Likelihood Meets Prediction-Powered Inference}
\author{Guanghui Wang, Mengtao Wen, Changliang Zou\thanks{Authors are listed in alphabetical order.}}
\affil{\it School of Statistics and Data Science, Nankai University}
\date{}

\begin{document}
\maketitle

\begin{bibunit}

\begin{abstract}
We study inference with a small labeled sample, a large unlabeled sample, and high-quality predictions from an external model. We link prediction-powered inference with empirical likelihood by stacking supervised estimating equations based on labeled outcomes with auxiliary moment conditions built from predictions, and then optimizing empirical likelihood under these joint constraints. The resulting empirical likelihood-based prediction-powered inference (EPI) estimator is asymptotically normal, has asymptotic variance no larger than the fully supervised estimator, and attains the semiparametric efficiency bound when the auxiliary functions span the predictable component of the supervised score. For hypothesis testing and confidence sets, empirical likelihood ratio statistics admit chi-squared-type limiting distributions. As a by-product, the empirical likelihood weights induce a calibrated empirical distribution that integrates supervised and prediction-based information, enabling estimation and uncertainty quantification for general functionals beyond parameters defined by estimating equations. We present two practical implementations: one based on basis expansions in the predictions and covariates, and one that learns an approximately optimal auxiliary function by cross-fitting. In simulations and applications, EPI reduces mean squared error and shortens confidence intervals while maintaining nominal coverage.
\end{abstract}

\section{Introduction}\label{sec:intro}

Across many fields, modern machine learning delivers high-quality predictions at far lower cost and far greater scale than collecting gold-standard labels \citep[e.g.,][]{AlphaFold2021,brown2020language}. As a result, empirical analyses often combine a small sample of observations with trusted outcomes and a much larger set of covariates and model-predicted values \citep{Wang2020pnas,Angelopoulos2023ppi}. The central question is how to use these predictions to improve efficiency while retaining valid inference for a well-defined population target.

We consider a parameter of interest \(\theta^*\in\bR^p\) defined as the solution to the estimating equation $\E\{g_{\theta}(Y, X)\} = 0$, where $Y\in\mathcal{Y}$ is the response, $X\in\mathcal{X}$ is a covariate vector, and \(g_{\theta}:\cY\times\cX\rightarrow\bR^p\) is an estimating function or score. Given only $n$ observations $\{(Y_i,X_i)\}_{i=1}^n$ with trusted labels, a standard supervised estimator solves $n^{-1}\sum_{i=1}^n g_{\theta}(Y_i,X_i)=0$ and is consistent under classical regularity conditions. In many modern applications, however, we also observe a much larger set of $m$ covariate vectors $\{X_i\}_{i=n+1}^{n+m}$, together with predictions $\tY_i=f(X_i)$ from a pre-trained model $f$. A naive imputation estimator that solves $m^{-1}\sum_{i=n+1}^{n+m} g_{\theta}(\tY_i,X_i)=0$ uses the abundant unlabeled data, but is generally biased unless $f$ predicts perfectly.

\textit{Prediction-powered inference} formalizes how to combine labeled and predicted data \citep{Angelopoulos2023ppi}. It combines the \textit{supervised score} \(g_{\theta}(Y, X)\) with an \textit{imputation contrast} that compares moments of \(g_{\theta}(\tY, X)\) between the labeled and unlabeled samples. Procedures in this family construct confidence sets that maintain nominal coverage and typically become shorter as prediction quality improves. Subsequent works \citep{angelopoulos2024ppi++,Gan2024PDC,miao2023assumption} emphasized two themes---\textit{safety} and \textit{efficiency}---by designing estimators whose asymptotic variance is never larger than that of the supervised estimator and is often substantially smaller. A key design choice is how to learn a \textit{power-tuning} parameter that balances supervised and imputed information.

Recently, \cite{Ji2025RePPI} unified many prediction-powered procedures as solutions to
\begin{equation}\label{eq:ppi}
n^{-1}\sum_{i=1}^n g_{\theta}(Y_i,X_i) - \Big\{n^{-1}\sum_{i=1}^n s_{\theta}(\tY_i,X_i)-m^{-1}\sum_{i=n+1}^{n+m} s_{\theta}(\tY_i,X_i)\Big\}=0, 
\end{equation}
for a suitable choice of \(s_\theta : \cY\times \cX \to \bR^p\). Many existing methods correspond to \(s_{\theta}(\tY, X) = M g_{\theta}(\tY, X)\) for some matrix \(M \in\bR^{p\times p}\).
In particular, this reduces to the original prediction-powered estimator when $M$ is the identity matrix \citep{Angelopoulos2023ppi}, and recovers power-tuned variants by optimizing $M$ over a pre-specified class \citep{angelopoulos2024ppi++,Gan2024PDC,miao2023assumption}.
In this view, an optimal choice of $s_\theta$ should satisfy $s_{\theta^*}(\tY,X) = (1-\gamma)\E\{g_{\theta^*}(Y,X)\mid \tY,X\}$, where $\gamma = \lim_{n,m\to\infty}n/(n+m) \in [0, 1]$ is the asymptotic proportion of labeled samples, and in practice $s_{\theta^*}$ is typically learned via flexible machine-learning methods combined with power tuning to ensure safety.

In this paper, we revisit prediction-powered inference through the lens of \textit{empirical likelihood}. 
Empirical likelihood assigns probability weights to observations so that a set of moment constraints hold, and uses the resulting likelihood for estimation and inference \citep{Owen1988,Owen2001EL,Qin1994EL}. 
We propose an \textit{empirical likelihood-based prediction-powered inference} framework, which we refer to as EPI. EPI uses the supervised score \(g_{\theta}(Y, X)\) together with an \textit{auxiliary contrast} \(h(\tY, X) - (n+m)^{-1}\sum_{i = 1}^{n+m} h(\tY_i, X_i)\) to form joint moment constraints in a single empirical likelihood problem, where \(h: \cY\times \cX \to \bR^r\) is an \textit{auxiliary function} that does not depend on \(\theta\). We view $h$ as a summary of $(\tY, X)$ that captures how the predictions are related to the supervised score.
The empirical likelihood optimization then automatically balances information from the supervised score and the auxiliary contrast, rather than relying on an explicit power-tuning parameter. 
We show that the resulting EPI estimator is safe: its asymptotic variance is no larger than that of the supervised estimator based solely on labeled data, for any fixed choice of \(h\). We further characterize an associated semiparametric efficiency bound. Let $s^*(\tY,X) = \E\{g_{\theta^*}(Y,X)\mid \tY,X\}$ denote the predictable component of the supervised score. When the centered auxiliary function \(h(\tY, X) - \E\{h(\tY, X)\}\) spans \(s^*(\tY,X)\), the EPI estimator attains the corresponding semiparametric efficiency bound. In this sense, EPI recovers the efficiency guarantees of optimally tuned prediction-powered estimators, while eliminating the need for an explicit power-tuning step.

We develop two practical implementations of auxiliary functions. The first uses basis expansions in \((\tY,X)\) to enlarge the span of \(h\) in a simple, deterministic way. The second learns \(s^*\) from the data and incorporates it through a cross-fitted auxiliary constraint, allowing flexible machine-learning methods while preserving validity. For both constructions, we derive asymptotic theory and construct confidence sets based on empirical likelihood ratio statistics.

Our main contributions are as follows:
\begin{itemize}
    \item \textbf{An empirical likelihood formulation of prediction-powered inference.} We introduce EPI, a framework that combines supervision-based scores and prediction-based auxiliary contrasts via a single empirical likelihood maximization. This formulation guarantees asymptotic safety relative to the supervised estimator and attains the semiparametric efficiency bound when the auxiliary construction spans the predictable component of the supervised score.
    \item \textbf{Prediction-powered inference with empirical likelihood advantages.} EPI inherits key advantages of empirical likelihood, including likelihood-ratio-type procedures that better adapt to asymmetry and other non-normal features, as well as a natural way to incorporate \textit{over-identified} scores arising from multiple valid moment conditions, as in generalized method of moments settings. 
    \item \textbf{Prediction-powered distribution learning.} The empirical likelihood weights define a calibrated empirical distribution that fuses labeled data and predictions. This enables estimators and confidence sets for general functionals, not only parameters defined by estimating equations.
\end{itemize}

\subsection{Related work}

A closely related line of work is \textit{semi-supervised estimation and inference}, which also leverages a small labeled sample together with a large unlabeled sample, but does not assume access to an external pre-trained model. Instead, a prediction function is learned from the observed data, typically using flexible regression or machine-learning tools. Existing results cover a range of targets, including prediction evaluation \citep{Gronsbell2018evaluate}, the mean of the response \citep{zhang2019semi,zhang2022high}, linear regression functionals such as coefficients and explained variance \citep{chakrabortty2018efficient,azriel2021semi,Deng2023semi,Cai2020ssvaraince}, and more general M-estimation and U-statistic problems \citep{Song2023,kim2024semi}. In addition, \citet{Wen2025semi} proposed a bias-amended semi-supervised procedure for distribution learning. Extending EPI to a fully semi-supervised regime is an interesting direction for future work.

Within the prediction-powered framework, most existing work assumes that a reasonably accurate pre-trained model is available, although there are extensions that first train the predictor from the observed data and then plug it into prediction-powered procedures (\citeauthor{Zrnic2024CPPI}, \citeyear{Zrnic2024CPPI}; cross-prediction-powered inference). These contributions focus primarily on finite-dimensional parameters; to our knowledge, distributional targets in the prediction-powered setting have not yet been studied, whereas our empirical likelihood approach provides a unified route to both parametric and distributional inference.

Empirical likelihood has a long history as a method for incorporating auxiliary information through moment conditions. A closely related contribution is \citet{Chen2003EL}, which considered \textit{surrogate endpoint} studies with a small validation sample containing both the primary outcome and a surrogate $(Y, S, X)$ and a large nonvalidation sample with only the surrogate and covariates $(S, X)$, and used a two-sample empirical likelihood to combine estimating equations from both samples for parameter inference. In their framework, both the primary and surrogate components are represented by zero-mean estimating functions specified a priori. In our prediction-powered setting, model predictions $\tY$ play a role analogous to surrogates, but we allow general auxiliary functions of $(\widetilde{Y}, X)$ that can be constructed or learned from the data and calibrated via centering and cross-fitting to yield valid empirical likelihood constraints. Conceptually, both approaches stack primary and auxiliary information within an empirical likelihood framework; we develop this perspective in the prediction-powered regime and use the induced weights for distributional as well as parametric inference. In this way, we bring classical empirical likelihood into contact with modern prediction-powered methodology; to our knowledge, empirical likelihood has not previously been linked explicitly to prediction-powered or semi-supervised procedures in this way.

\subsection{Organization}

The rest of the paper is organized as follows. Section~\ref{sec:method} formally introduces the proposed empirical likelihood methods for prediction-powered inference, including ideal and finite-unlabeled regimes and the construction of auxiliary functions. Section~\ref{sec:extension} presents extensions to over-identified supervised scores and prediction-powered distribution learning. Simulation results and a real-data application are reported in Section~\ref{sec:simu}. We conclude in Section~\ref{sec:con}.

\textbf{Notation.} For a measurable function \(f(Y,\tY,X)\), we write \(\E(f) \equiv \E\{f(Y, \tY, X)\}\) and \(\cov(f) \equiv \cov\{f(Y, \tY, X)\}\) for short. For example, \(\E(g_\theta) = \E\{g_\theta(Y, X)\}\) and \(\E(h) = \E\{h(\tY, X)\}\). For a function \(f_\theta\) indexed by \(\theta\), let \(\partial f_{\theta}/\partial \theta\) and \(\partial^2 f_\theta/\partial \theta\partial\theta^\T\) denote its first- and second-order derivatives with respect to \(\theta\), respectively. For a vector \(v\), \(\|v\|\) denotes its Euclidean norm, and for a matrix \(A\), \(\|A\|\) denotes its operator norm. For two square matrices \(M_1, M_2 \in \bR^{d\times d}\), we write \(M_1 \preceq M_2\) (equivalently, \(M_2 \succeq M_1\)) if \(M_2 - M_1\) is positive semidefinite.

\section{Empirical likelihood for prediction-powered inference}\label{sec:method}

\subsection{An empirical likelihood perspective}\label{sec:method_ideal}

We begin with an ideal setting in which the distribution of $X$ is known, corresponding to \(m=\infty\). In this regime, population moments involving \((\tY, X)\) are available. This allows us to highlight how empirical likelihood can incorporate prediction-based auxiliary information before accounting for additional variability from estimating moments of \((\tY, X)\).

Our target is \(\theta^*\), defined by the solution to the estimating equation
\[
    \E\{g_{\theta}(Y, X)\} = 0.
\]
\begin{exmp}\label{exmp:mean}
    For the mean of the response $\theta^*=\E(Y)$, take $g_{\theta}(Y, X)=Y-\theta$. 
\end{exmp}
\begin{exmp}\label{exmp:lm}
    For the coefficients of the linear projection $\theta^*=\{\E(XX^\top)\}^{-1}\E(XY)$, take $g_{\theta}(Y, X)=X(X^\T\theta - Y)$. 
\end{exmp}

In prediction-powered settings, we also introduce an \(r\)-dimensional auxiliary function \(h\). We view $h$ as a summary of \((\tY, X)\) that captures predictive information relevant to the supervised score $g_{\theta}(Y, X)$, and may therefore improve inference for \(\theta^*\). For mean inference in Example~\ref{exmp:mean}, a simple choice is \(h(\tY, X) = \tY - \E(\tY)\), where \(\E(\tY)\) is known in this ideal setting. Without loss of generality, assume \(\E\{h(\tY, X)\} = 0\).

Empirical likelihood provides a likelihood-style framework for combining multiple moment conditions without specifying a parametric model, making it well suited to integrating supervised and prediction-based information. Assign probability weights $w_1,\ldots,w_n$ to the labeled observations and define the \textit{profile empirical likelihood} 
\begin{equation}\label{eq:h-ideal}
    \begin{split}
        \ell_n(\theta;h):=\max\Big\{\prod_{i=1}^n w_i: ~&w_i\geq 0,\ \sum_{i=1}^n w_i=1,\ \sum_{i=1}^n w_i g_{\theta}(Y_i,X_i) = 0,\\
        &\sum_{i=1}^n w_i h(\tY_i,X_i) = 0\Big\}.  
    \end{split}
\end{equation}
We refer to
\[
    \hat{\theta}_{\EL; h} := \underset{\theta}{\arg\max}\ \ell_n(\theta;h)
\]
as the \textit{empirical likelihood-based prediction-powered inference} (\textit{EPI}) estimator.

\begin{remark}[On computation]
The empirical likelihood problem based on the combined functions \(g_\theta\) and \(h\) falls within standard empirical likelihood theory for multiple estimating functions \citep[e.g.,][]{Qin1994EL}, with the special structure that \(h\) does not depend on \(\theta\). This structure allows predictive information to enter through the empirical likelihood weights, while the estimation of \(\theta\) remains a weighted estimating equation based on \(g_\theta\). Specifically, \(\hat{\theta}_{\EL; h}\) can be computed by profiling. First find weights satisfying only the auxiliary constraint
\begin{equation}\label{eq:weights}
    (\hat{w}_1,\ldots,\hat{w}_n) := \arg\max\Big\{\prod_{i = 1}^n w_i:\ w_i\geq 0,\ \sum_{i = 1}^n w_i = 1,\ \sum_{i = 1}^n w_i h(\tY_i, X_i) = 0\Big\}.
\end{equation}
We then estimate $\theta$ by solving
\[
\sum_{i = 1}^n \hat{w}_i g_\theta(Y_i, X_i) = 0.
\]
Intuitively, the auxiliary constraint determines the weights and the supervised estimating equation selects the value of \(\theta\) compatible with those weights, which yields the same \(\hat{\theta}_{\EL; h}\) as maximizing \eqref{eq:h-ideal}. In practice, both steps can be implemented using standard solvers.
\end{remark}

We next provide large-sample guarantees for \(\hat{\theta}_{\EL; h}\) in this ideal setting.

\begin{asmp}\label{asmp:EL}
    (a) \(J:=\E\{\partial g_{\theta^*}/\partial\theta\}\) has rank \(p\).
    (b) There exists a neighborhood \(\Theta\) of \(\theta^*\) such that \(\partial g_\theta(y, x)/\partial\theta\) and \(\partial^2 g_\theta(y, x)/\partial \theta\partial\theta^\T\) are continuous for all \(\theta\in\Theta\), and for an integrable function \(G(y, x)\), \(\max\{\|\partial g_\theta(y, x)/\partial\theta\|, \|g_\theta(y, x)\|^3, \|\partial^2 g_\theta(y, x)/\partial \theta\partial\theta^\T\|\} \leq G(y, x)\) for all \(\theta\in\Theta\).
\end{asmp}

\begin{prop}\label{thm:ideal}
    Suppose Assumption~\ref{asmp:EL} holds. If \(\cov\{(g_{\theta^*}^\T, h^\T)^\T\}\) is positive definite and \(\E\{\|h\|^3\} < \infty\), then 
    \begin{align*}
        \sqrt{n}(\hat{\theta}_{\EL; h} - \theta^*) \rightsquigarrow \mathcal{N}(0,\Sigma_{h}),
    \end{align*}
    where
    $\Sigma_{h} = J^{-1} \left\{\cov(g_{\theta^*}) - \E(g_{\theta^*}h^\T)\cov(h)^{-1} \E(hg_{\theta^*}^\T)\right\} J^{-\T}.
    $
\end{prop}

Assumption~\ref{asmp:EL} is standard in the empirical likelihood literature \citep[e.g.,][]{Qin1994EL,Owen2001EL}. The adjustment term $\E(g_{\theta^*}h^\T)\cov(h)^{-1} \E(hg_{\theta^*}^\T)$ in \(\Sigma_{h}\) is positive semidefinite, implying
\[
    \Sigma_{h} \preceq J^{-1} \cov(g_{\theta^*}) J^{-\T},
\]
where $J^{-1} \cov(g_{\theta^*}) J^{-\T}$ is the asymptotic covariance of the supervised estimator based only on $g_\theta(Y, X)$. Thus, incorporating predictive information through \(h\) is safe in this ideal setting.

We can further interpret the role of \(h\) through an efficiency bound for this ideal prediction-powered model in which the distribution of \(X\) is known. Define
\[
    s^*\equiv s^*(\tY, X) := \E\{g_{\theta^*}(Y, X)\mid \tY, X\},
\]
the predictable component of the supervised score given \((\tY,X)\). A standard projection argument implies that for any auxiliary function \(h\), 
\[
    \E(g_{\theta^*}h^\T)\cov(h)^{-1} \E(hg_{\theta^*}^\T) \preceq \cov(s^*).
\]
Consequently,
\[
    \Sigma_{h} \succeq J^{-1} \{\cov(g_{\theta^*}) - \cov(s^*)\}J^{-\T},
\]
with equality when there exists a matrix \(A\in\bR^{p\times r}\) such that \(s^* = A h\). When \(P_X\) is unknown and \(\E\{h(\tY,X)\}\) must be estimated from a finite unlabeled sample, the efficiency bound has the same form with a modified variance reduction term that accounts for this additional estimation step; see Section~\ref{sec:method_ss}.

Empirical likelihood also yields a likelihood-ratio inference procedure for \(\theta^*\), providing an alternative to Wald-type inference commonly used in prediction-powered settings.

\begin{prop}\label{thm:LR-ideal}
    Under the conditions of Proposition~\ref{thm:ideal},
    \[
        T_n (\theta^*; h) := -2\log\{\ell_n(\theta^*; {h})/\ell_n(\hat{\theta}_{\EL; h}; {h})\} \rightsquigarrow \chi^2_p.
    \] 
\end{prop}

An empirical likelihood ratio confidence set for $\theta^*$ is therefore
\[
    \{\theta: T_n(\theta; h) \leq \chi^2_{p, 1-\alpha}\},
\]
where \(\chi^2_{p, 1-\alpha}\) is the \((1-\alpha)\)-quantile of the chi-squared distribution with \(p\) degrees of freedom. This Wilks-type result delivers likelihood-ratio tests and confidence sets without explicit variance estimation.

The proofs of Propositions~\ref{thm:ideal}--\ref{thm:LR-ideal} follow from standard empirical likelihood results for multiple estimating functions \citep{Qin1994EL}. With \(h\) free of \(\theta\), our asymptotic normality and Wilks-type conclusions agree with the corresponding results in \citet{Qin1994EL}.

\subsection{Auxiliary functions with unknown means}\label{sec:method_ss}

We now consider the practical setting with a finite unlabeled sample \(m < \infty\). In this regime, the moment \(\E\{h(\tY, X)\}\) is generally unknown because \((\tY, X)\) is observed only through the \(n\) labeled and \(m\) unlabeled observations. We therefore center \(h\) by its pooled sample mean and impose the centered auxiliary moment in the empirical likelihood constraints.

Define the centered auxiliary function as
\[
    h(\tY, X) - \frac{1}{n+m}\sum_{i=1}^{n+m} h(\tY_i, X_i),
\]
and set
\begin{equation}\label{eq:h-center}
    \begin{split}
        \ell_n(\theta;h):=\max\Big\{\prod_{i=1}^n w_i: ~&w_i\geq 0,\ \sum_{i=1}^n w_i=1,\ \sum_{i=1}^n w_i g_{\theta}(Y_i,X_i) = 0,\\
        &\sum_{i=1}^n w_i \big\{h(\tY_i,X_i) - \frac{1}{n+m}\sum_{j=1}^{n+m} h(\tY_j,X_j)\big\}= 0\Big\}.  
    \end{split}
\end{equation}
Maximizing \(\ell_n(\theta; h)\) over \(\theta\) yields the corresponding estimator \(\hat{\theta}_{\EL; h}\). The two-step computation in Section~\ref{sec:method_ideal} remains applicable, with the auxiliary constraint in \eqref{eq:weights} replaced by the centered version in \eqref{eq:h-center}.

\begin{remark}[Empirical likelihood analogue of imputation contrasts]
Many prediction-powered procedures can be viewed as solutions to a rectified estimating equation involving a suitable choice of \(s_\theta(\tY,X)\) \citep{Ji2025RePPI}; see also Eq.~\eqref{eq:ppi}. The centered auxiliary constraint in \eqref{eq:h-center} provides an empirical likelihood analogue of this imputation contrast. Our formulation differs in that \(h\) is free of \(\theta\). We introduce a summary \(h(\tY,X)\), and let the empirical likelihood weights combine supervised and auxiliary moments adaptively for the target \(\theta^*\).
\end{remark}

We study the large-sample properties of \(\hat{\theta}_{\EL; h}\) under this centered formulation. Assume \(n/(n+m)\to \gamma\in[0,1]\), where \(\gamma\) is the asymptotic labeled proportion in the pooled sample.

\begin{thm}\label{thm:ss}
    Suppose Assumption~\ref{asmp:EL} holds. If \(\cov((g_{\theta^*}^\T, h^\T)^\T)\) is positive definite and \(\E\{\|h\|^3\} < \infty\), then
    \begin{align*}
        \sqrt{n}(\hat{\theta}_{\EL; h}-\theta^*) \rightsquigarrow \mathcal{N}(0,\Sigma_{h}),
    \end{align*}
    where 
    \[
        \Sigma_{h} = J^{-1} \left\{\cov(g_{\theta^*}) - (1-\gamma) \E(g_{\theta^*}h^\T)\cov(h)^{-1}\E(h g_{\theta^*}^\T)\right\} J^{-\T}.
    \]
\end{thm}

Relative to Proposition~\ref{thm:ideal}, the prediction-based adjustment is multiplied by \((1-\gamma)\), reflecting the additional uncertainty from estimating \(\E\{h(\tY,X)\}\) using the pooled sample. When \(\gamma = 0\), Theorem~\ref{thm:ss} reduces to Proposition~\ref{thm:ideal}.

The adjustment term remains positive semidefinite, preserving the safety guarantee. Moreover, efficiency continues to be governed by how well the centered auxiliary function captures \(s^*(\tY,X)=\E\{g_{\theta^*}(Y,X)\mid \tY,X\}\). In particular, efficiency is attained when there exists \(A\in\mathbb{R}^{p\times r}\) such that \(s^*(\tY,X)=A[h(\tY,X)-\E\{ h(\tY,X)\}]\), where the semiparametric efficiency bound takes the ideal form with the variance reduction term scaled by \((1-\gamma)\), consistent with related prediction-powered efficiency characterizations \citep{Ji2025RePPI}.

\begin{thm}\label{thm:LR-ss}
    Under the conditions of Theorem~\ref{thm:ss}, 
    \[
        T_n(\theta^*;h) := -2\log\{\ell_n(\theta^*; h)/\ell_n(\hat{\theta}_{\EL; h}; h)\} \rightsquigarrow \sum_{j = 1}^p \lambda_j \xi_j^2,
    \]
    where \(\xi_1, \ldots, \xi_p\) are independent standard normal random variables, and \(\lambda_1, \ldots, \lambda_p\) are eigenvalues of 
    \[
        \Lambda_h = I_p + \gamma (J^\T U_h^{-1}J)^{1/2} J^{-1} \E(g_{\theta^*} h^\T) \{\cov(h)\}^{-1} \E(h g_{\theta^*}^\T) J^{-\T} (J^\T U_h^{-1} J)^{1/2},
    \]
    with \(U_h = \cov(g_{\theta^*}) - \E(g_{\theta^*}h^\T) \{\cov(h)\}^{-1} \E(hg_{\theta^*}^\T)\).
\end{thm}

Theorem~\ref{thm:LR-ss} shows that the empirical likelihood ratio no longer has a standard chi-squared limit because \(\E(h)\) is estimated (via centering), consistent with empirical likelihood theory when nuisance quantities enter the constraints \citep[e.g.,][]{Hjort2009EL}. We adopt this centering strategy because it preserves the computational advantages of profiling while accounting for uncertainty in estimating \(\E(h)\). For inference, we estimate \(\lambda_1, \ldots, \lambda_p\) using plug-in estimators of \(\Lambda_h\), compute the corresponding upper quantile \(t_\alpha\), and form the confidence set
\[
    \{\theta: T_{n}(\theta; h) \leq t_\alpha\},
\]
with implementation details given in Section~\ref{supp:simu} of the supplementary material.

\subsection{Designing auxiliary functions}

Section \ref{sec:method_ss} shows that efficiency gains arise when the centered auxiliary function \(h(\tY,X)-\E\{h(\tY,X)\}\) can approximate
\[
s^*(\tY,X) = \E\{g_{\theta^*}(Y,X)\mid \tY,X\},
\]
the predictable component of the supervised score given \((\tY,X)\). In general, \(s^*\) is unknown because it depends on the target \(\theta^*\) and on this conditional expectation. We consider two complementary strategies for constructing \(h\): basis expansions that enlarge the auxiliary span through predetermined features of \((\tY,X)\), and cross-fitted constructions that learn \(s^*\) from the labeled data.

\subsubsection{Basis expansions for auxiliary functions}\label{sec:method:basis}

The basis-expansion strategy specifies \(h\) as a vector of predetermined basis functions evaluated at \((\tY,X)\). This approach aligns with a projection theme common in semi-supervised inference: enrich a dictionary of covariate-based summaries to improve the representation of prediction-relevant structure \citep{chakrabortty2018efficient,zhang2019semi,azriel2021semi,Song2023}. Here it provides a simple and easily implementable way to enlarge the span of \(h\).

We illustrate this idea using the linear projection target $\theta^*=\{\E(XX^\top)\}^{-1}\E(XY)$, with the supervised score $g_{\theta}(Y, X)=X(X^\T\theta - Y)$. Then
\[
    s^*(\tY, X) = X\{X^\T\theta^* - \E(Y\mid \tY, X)\}.
\]
When the pre-trained predictor is accurate so that \(\tY\) closely approximates \(\E(Y\mid X)\), it is natural to expect \(\E(Y\mid \tY,X)\) to be close to \(\tY\), and hence
\[
    s^*(\tY,X)\approx X(X^\T\theta^*-\tY)=\sum_{k=1}^p \theta_k^*\, X X_k \;-\; X\tY,
\]
where, for a vector $a$, \(a_k\) denotes its \(k\)th component. This approximation lies in the linear span of interaction terms \(X_jX_k\) and \(X_j\tY\), motivating
\[
    h(\tY, X) = (X_1^2, X_1X_2, \ldots, X_p^2, X_1\tY, \ldots, X_p\tY)^\T.
\]
With this choice, the span of \(h-\E(h)\) can represent a dominant part of \(s^*\) even though \(\theta^*\) is unknown.

When prediction quality is uncertain, \(h\) can be enriched with higher-order polynomials, splines, or other dictionary terms in \((\tY,X)\). Theorem \ref{thm:ss} implies that such basis-expanded choices preserve safety relative to the supervised benchmark. In addition, adding further valid basis components cannot worsen asymptotic variance; see Corollary~1 of \citet{Qin1994EL}. A richer basis can therefore improve the approximation of \(s^*\) and enhance efficiency in practice. Once a basis is specified, the empirical likelihood formulation in \eqref{eq:h-center} applies directly.

\subsubsection{Cross-fitted auxiliary functions}\label{sec:method_crossfit}

An alternative to basis expansions is to construct \(h\) by learning $s^*(\tY,X)=\E\{g_{\theta^*}(Y,X)\mid \tY,X\}$. A naive plug-in approach may, however, invalidate the auxiliary moment constraint because the learned auxiliary function would be evaluated on the same observations that impose the constraint. We therefore adopt cross-fitting \citep{chernozhukov2018double} so that the auxiliary constraint remains asymptotically valid while allowing \(h\) to approximate \(s^*\).

Learning \(s^*(\tY,X)\) is also central to recent prediction-powered developments. For example, \citet{Ji2025RePPI} targets the same \(s^*(\tY,X)\) and adopts a three-fold cross-fitting scheme: separate folds are used to obtain an initial estimate of \(\theta^*\), to estimate the conditional expectation defining \(s^*\), and to carry out prediction-powered inference with power tuning; the roles of these folds are then rotated and the resulting estimators are averaged. Our approach incorporates the learned \(s^*\) through a cross-fitted auxiliary constraint within a single empirical likelihood maximization.

Assume that \(K\) divides both \(n\) and \(m\). Randomly partition the labeled indices \(\{1,\ldots,n\}\) into \(K\) folds \(\cI_1,\ldots,\cI_K\) and the unlabeled indices \(\{n+1,\ldots,n+m\}\) into \(K\) folds \(\cJ_1,\ldots,\cJ_K\), with sizes \(n_K=n/K\) and \(m_K=m/K\). For each \(k\), learn $s^*$ using labeled data indexed by \(\{1,\ldots,n\}\setminus \cI_k\) to obtain \(\hat h^{(-k)}\). For \(i\in\cI_k\), define the fold-specific centered auxiliary term 
\[
    \hat{h}^{(-k)}(\tY_i,X_i) - \frac{1}{n_K+m_K}\sum_{\ell \in \cI_k\cup\cJ_k} \hat{h}^{(-k)}(\tY_\ell, X_\ell).
\]
Combining these constraints across \(k=1,\ldots,K\), define the cross-fitted profile empirical likelihood
\begin{equation}\label{eq:h-crossfit}
    \begin{split}
        \ell_n(\theta):=\max\Big\{\prod_{i=1}^n w_i: ~&w_i\geq 0, \sum_{i=1}^n w_i=1,\sum_{i=1}^n w_i g_{\theta}(Y_i,X_i) = 0,\\
        &\sum_{k=1}^K\sum_{i \in\cI_k} w_i \big\{\hat{h}^{(-k)}(\tY_i, X_i) - \frac{1}{n_K+m_K}\sum_{\ell \in \cI_k\cup\cJ_k} \hat{h}^{(-k)}(\tY_\ell, X_\ell)\big\} = 0\Big\}.
    \end{split}
\end{equation}
Maximizing \(\ell_n(\theta)\) over \(\theta\) yields the corresponding estimator \(\hat{\theta}_{\EL}\). Computation follows the profiling workflow in Section~\ref{sec:method_ss} with the auxiliary constraint replaced by its cross-fitted version. 

\begin{asmp}\label{asmp:learn-h}
There exists a nonrandom function \(\h(\tY,X)\) such that, \[\max_{k=1,\ldots,K}\E[\|\hat{h}^{(-k)}(\tY, X) - \h(\tY, X)\|^2\mid \hat{h}^{(-k)}] = o_P(1).\]
\end{asmp}

\begin{thm}\label{thm:crossfit}
Suppose Assumptions~\ref{asmp:EL}--\ref{asmp:learn-h} hold. If \(\cov((g_{\theta^*}^\T, \h^\T)^\T)\) is positive definite and \(\E\{\|\h\|^3\} < \infty\), then the asymptotic normality in Theorem~\ref{thm:ss} and the empirical likelihood ratio limit in Theorem~\ref{thm:LR-ss} remain valid, with \(h\) replaced by \(\h\).
\end{thm}

In particular, \(\hat{\theta}_{\EL}\) remains asymptotically safe relative to the supervised benchmark. Moreover, when the span of \(\h-\E(\h)\) contains \(s^*\), \(\hat{\theta}_{\EL}\) attains the corresponding prediction-powered efficiency bound.

\begin{remark}[Basis versus cross-fitted auxiliary designs]
The basis-expansion and cross-fitted constructions provide complementary ways to approximate \(s^*(\tY,X)\), and both preserve the safety guarantees established above. Basis expansions rely on a pre-specified dictionary of \((\tY,X)\) features; they are simple to implement and particularly appealing when a low-dimensional representation of \(s^*\) is plausible. Cross-fitted constructions instead learn \(s^*\) from the data and then embed the learned function into EPI via a cross-fitted auxiliary constraint, offering greater flexibility in complex or high-dimensional problems at the cost of additional computation.
\end{remark}

\section{Extensions: over-identified estimating equations and distribution learning}\label{sec:extension}

\subsection{Over-identified estimating equations}

We begin by clarifying identification for the supervised estimating equation. Let \(g_\theta:\mathcal Y\times\mathcal X\to\mathbb R^q\) satisfy \(\E\{g_{\theta^*}(Y,X)\}=0\). When \(q=p\) and the Jacobian \(J=\E(\partial g_{\theta^*}/\partial\theta)\) has full rank \(p\), we refer to \(g_\theta\) as \emph{just-identified}. When \(q>p\), \(g_\theta\) is \emph{over-identified}, meaning that \(\theta^*\) is characterized by more moment conditions than parameters. Empirical likelihood is particularly appealing in this regime because it accommodates multiple moment restrictions without requiring an explicit choice of an optimal weighting matrix, in contrast to generalized method of moments procedures \citep{Hansen1982}.

Throughout this subsection, we replace the \(p\)-dimensional supervised score in Section~\ref{sec:method} by the \(q\)-dimensional function \(g_\theta\) above. The profile empirical likelihoods in the ideal case \eqref{eq:h-ideal}, the centered case \eqref{eq:h-center}, and the cross-fitted case \eqref{eq:h-crossfit} remain well defined. We continue to denote the resulting estimators by \(\hat\theta_{\EL; h}\) for formulations that explicitly use \(h\), and by \(\hat\theta_{\EL}\) for the cross-fitted construction.

\begin{remark}[On computation]
The profiling scheme in Section~\ref{sec:method_ideal} exploits the \(\theta\)-free structure of \(h\) and the fact that the supervised constraint is \(p\)-dimensional. With an over-identified \(g_\theta\), the constraint \(\sum_{i=1}^n w_i g_\theta(Y_i,X_i)=0\) is \(q\)-dimensional. Imposing it jointly with the auxiliary constraint typically makes the feasible weights depend on \(\theta\), so solving \eqref{eq:weights} first and then inserting the resulting weights into a weighted supervised estimating equation does not, in general, recover the maximizer of the full profile empirical likelihood. We therefore compute \(\hat\theta_{\EL; h}\) and \(\hat\theta_{\EL}\) using standard empirical likelihood Lagrange-multiplier algorithms; see Section~\ref{supp:simu} of the supplementary material.
\end{remark}

We next state the asymptotic normality result for the centered formulation with an over-identified supervised score. The ideal regime in Section~\ref{sec:method_ideal} corresponds to \(\gamma=0\), and the cross-fitted analogue follows by replacing \(h\) with \(\h\). For brevity, we defer the likelihood-ratio theory for over-identified \(g_\theta\) to Theorem~\ref{THM:SUPP:LR} of the supplementary material.

\begin{thm}\label{thm:over}
    Under the conditions of Theorem~\ref{thm:ss},
    \begin{align*}
        \sqrt{n}(\hat{\theta}_{\EL; h}-\theta^*) \rightsquigarrow \mathcal{N}(0,\Sigma_{h}),
    \end{align*}
    where
    \begin{equation*}
        \Sigma_{h} = \left[J^{\T}U_h^{-1}J \left\{J^{\T}U_h^{-1}J + \gamma J^{\T}U_h^{-1} \E(g_{\theta^*} h^\T) \{\cov(h)\}^{-1} \E(hg_{\theta^*}^\T) U_h^{-1}J\right\}^{-1} J^{\T}U_h^{-1}J\right]^{-1},
    \end{equation*}
    and \(U_h = \cov(g_{\theta^*}) - \E(g_{\theta^*}h^\T) \{\cov(h)\}^{-1} \E(hg_{\theta^*}^\T)\).
\end{thm}

Theorem~\ref{thm:over} reduces to Theorem~\ref{thm:ss} when \(q=p\), in which case \(J\) is square and invertible.

When \(\gamma=0\), the additional variability from estimating \(\E\{h(\tY,X)\}\) becomes negligible. In this regime, Theorem~\ref{thm:over} aligns with the over-identified empirical likelihood covariance formulas in \citet{Qin1994EL}. The role of \(h\) continues to be governed by the predictable component of the supervised score \(s^*(\tY,X)=\E\{g_{\theta^*}(Y,X)\mid \tY,X\}\). Efficiency is achieved when the span of the centered \(h\) (or \(\h\) in the cross-fitted construction) contains \(s^*\). Accordingly, the basis-expansion and cross-fitted strategies in Section~\ref{sec:method:basis}--\ref{sec:method_crossfit} extend directly to this setting by targeting the \(q\)-dimensional \(s^*\).

When \(\gamma>0\), the interaction between over-identification and the estimated centering step can complicate a uniform safety comparison with the supervised benchmark. A sufficient condition ensuring safety in this regime is given in Theorem \ref{THM:SUPP:NORM} of the supplementary material. In practice, \(s^*(\tY,X)\) remains a natural target for constructing \(h\), though formal efficiency guarantees in the over-identified finite-\(m\) regime are left for future work.

\subsection{Prediction-powered distribution learning}\label{sec:extension_dist}

Beyond parameter inference, empirical likelihood also provides a prediction-powered route to learning the joint distribution of \((Y,X)\). We first consider a baseline construction in which the empirical likelihood weights are determined solely by auxiliary moment restrictions. This yields a calibrated empirical distribution that leverages predictive information without requiring a supervised target \(g_\theta\).

Let \(\hat{w}_1,\ldots,\hat{w}_n\) denote the empirical likelihood weights obtained from \eqref{eq:weights} in the ideal regime or its centered (cross-fitted) analogue in the finite-\(m\) regime. Define the calibrated empirical distribution function
\[
\hat{F}_{\EL}(y, x) = \sum_{i = 1}^n \hat{w}_i \1(Y_i\leq y, X_i\leq x),
\]
and let \(F(y,x)=\P(Y\le y, X\le x)\).

\begin{thm}\label{thm:dist}
    If \(\cov(h)\) is positive definite and \(\E\{\|h\|^3\} < \infty\), then for any \((y, x)\in \cY\times \cX\),
    \[
    \sqrt{n}\big\{\hat{F}_{\EL}(y, x) - F(y, x)\big\} \rightsquigarrow \mathcal{N}\left(0, \sigma^2_{\EL}(y, x)\right),
    \] 
    where 
    \[\begin{split}
        \sigma^2_{\EL}(y, x) &= F(y, x)\{1-F(y, x)\} - (1-\gamma)\cov\{h(\tY, X), \1(Y\leq y, X\leq x)\}^\T\\ 
        &\hspace{3cm}  [\cov\{h(\tY, X)\}]^{-1} \cov\{h(\tY, X), \1(Y\leq y, X\leq x)\}.
    \end{split}\]
\end{thm}

Theorem~\ref{thm:dist} implies that \(\hat{F}_{\EL}(y,x)\) improves upon the supervised empirical distribution estimator \(\hat{F}_{\cls}(y, x) = n^{-1}\sum_{i = 1}^n \1(Y_i\leq y, X_i\leq x)\) in pointwise asymptotic variance, since \(\sigma^2_{\EL}(y, x) \leq F(y, x)\{1-F(y, x)\}\) for all \((y,x)\). Thus, auxiliary moments yield a safe variance reduction for distribution learning.

The efficiency interpretation parallels Section~\ref{sec:method}. For estimating \(F(y,x)\) at a fixed point, the analogue of the predictable component is
\[
s^*_{F(y, x)}(\tY, X) := \E\{\1(Y\leq y, X\leq x)\mid \tY, X\}.
\]
The lower bound induced by this target takes the form
\[
\sigma^2_{\EL}(y, x) \geq F(y, x)\{1-F(y, x)\} - (1-\gamma)\cov\{s^*_{F(y, x)}(\tY, X)\},
\]
with equality when the span of \(h-\E(h)\) contains \(s^*_{F(y,x)}\). Accordingly, the basis-expansion and cross-fitted strategies in Section~\ref{sec:method:basis}--\ref{sec:method_crossfit} can be used to approximate \(s^*_{F(y,x)}\). When interest centers on the marginal distribution of \(Y\), i.e., \(x=(\infty,\ldots,\infty)^\T\), the optimal pointwise target becomes
\[
s^*_{F(y)}(\tY, X) := \E\{\1(Y\leq y)\mid \tY, X\},
\]
aligning with related semi-supervised distribution-learning results \citep{Wen2025semi}.

Finally, when an over-identified supervised score \(g_\theta\) is available, one may further incorporate these additional supervised moment restrictions into the empirical likelihood weights. This can yield additional efficiency gains for distribution learning beyond the auxiliary-only construction above; corresponding theory is provided in Theorem~\ref{THM:SUPP:DIST} of the supplementary material.

\section{Numerical studies}\label{sec:simu}

\subsection{Mean inference}\label{sec:simu:mean}

We first consider the mean inference problem in Example~\ref{exmp:mean}. 
Data are generated according to \(Y = X + \varepsilon\) and \(\tY = X + \tilde{\varepsilon}\), where \(X\) and \(\varepsilon\) are independent with \(X,\varepsilon \sim \cP\) and \(\tilde{\varepsilon}\sim\cN(0, 1)\) is independent of \((X,\varepsilon)\). 
We adopt the following benchmarks: the supervised ($\cls$) estimator, the prediction-powered inference (PPI) estimator \citep{Angelopoulos2023ppi}, the prediction-decorrelated (PDC) estimator \citep{Gan2024PDC}, and the recalibrated prediction-powered inference (RePPI) estimator \citep{Ji2025RePPI}.
For the proposed method, we consider the basis-expansion EPI (with \(h(\tY, X) = (\tY, X)^\T\))
and the cross-fitted EPI (using XGBoost as the learning algorithm). 
To evaluate estimation accuracy, we report mean squared error (MSE). For inference, we report the miscoverage probability and the average confidence-interval length. We present MSE and average length as ratios relative to those of the supervised estimator \(\hat{\theta}_{\cls}\); ratios below 1 indicates higher efficiency.
Furthermore, to assess how well the confidence intervals adapt to the underlying data distribution, we distinguish between cases where the true parameter \(\theta^*\) falls below the lower endpoint or above the upper endpoint of the interval, termed lower and upper miscoverage, respectively.
All results are averaged over 10000 replications.

\begin{figure}[!h]
    \centering
    \includegraphics[width=\textwidth]{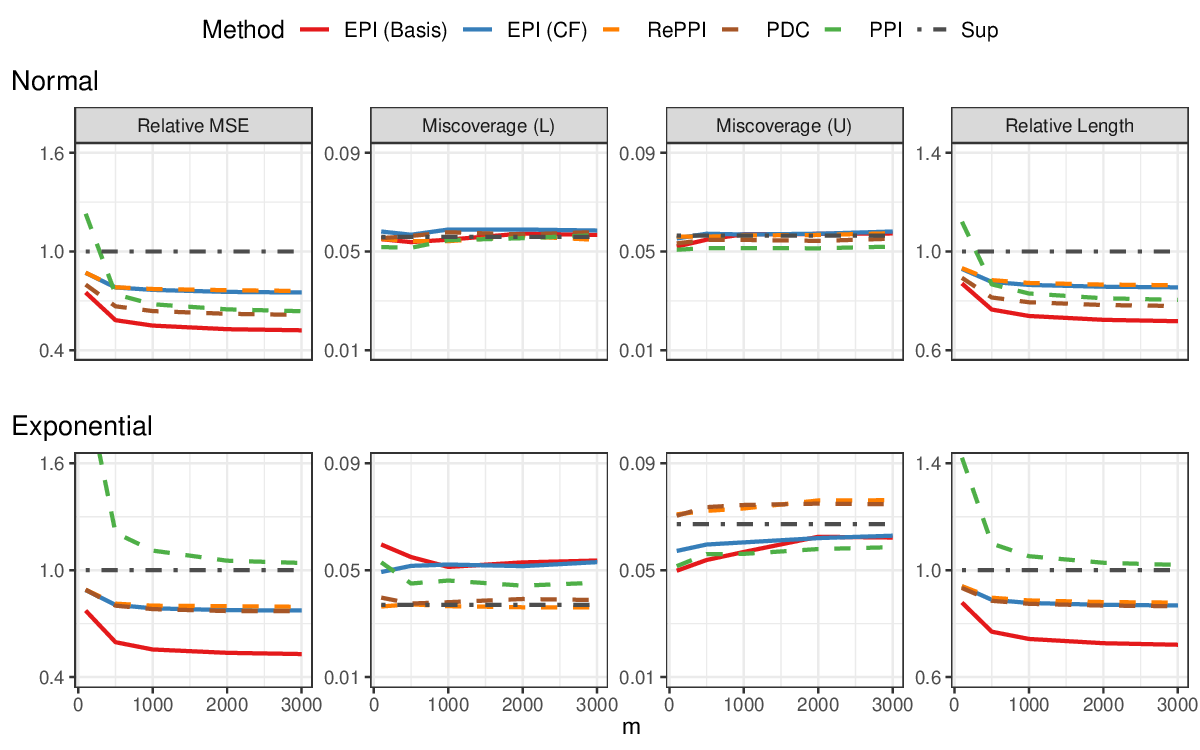}
    \caption{\small Performance of different methods in the mean inference problem.
    Both the mean squared error and the confidence-interval length are displayed as ratios relative to \(\hat{\theta}_{\cls}\). ``L'' and ``U'' denote lower and upper miscoverage, respectively.
    The nominal coverage level is \(1-\alpha=0.9\).
    EPI (Basis), the proposed basis-expansion estimator; EPI (CF), the proposed cross-fitted estimator; RePPI, the recalibrated prediction-powered inference; PDC, the prediction-decorrelated estimator; PPI, the prediction-powered inference; $\cls$, the supervised estimator.
    }
    \label{fig:mean}
\end{figure}

Figure~\ref{fig:mean} shows the results when \(n = 100\), where we vary \(m\in\{100, 500, 1000, 2000, 3000\}\) and \(\cP\in \{\cN(0, 1), \mathrm{Exp}(2)\}\). 
We first discuss efficiency in terms of relative MSE and relative interval length.
The results show that our \EPI estimators and the power-tuned variants of PPI remain safe relative to the supervised estimator, except for the original PPI, consistent with the theory.
The basis-expansion EPI performs best, since \(s^*(\tY, X) = X - \theta^*\) lies in the span of the centered basis \(h - \E(h)\) in this setting, while the cross-fitted EPI and RePPI incur finite-sample efficiency loss due to uncertainty in learning \(s^*\). 
Without the learning step, PDC outperforms RePPI and the cross-fitted EPI when the prediction quality of \(\tY\) is high (in the normal cases), but is inferior to the basis-expansion EPI because PDC only uses information from \(\tY\).
In terms of miscoverage, the total miscoverage probability (lower plus upper) of all methods is close to the nominal level \(\alpha=0.1\).
However, under the skewed setting (the exponential distribution), the imbalance between upper and lower miscoverage for the power-tuned PPI estimators becomes noticeable. In contrast, our EPI estimators adapt to distributional asymmetry, benefiting from the empirical likelihood framework.

\subsection{Linear regression}\label{sec:simu:lm}

Next, we consider inference for the coefficients in linear regression in Example~\ref{exmp:lm}. 
We generate data according to \(Y = \sum_{j = 1}^2 (X_j + X_j^2) + \varepsilon/2\) and \(\tY = \sum_{j = 1}^2 (X_j + X_j^2) + \rho \tilde{\varepsilon}\), where \(\varepsilon\sim t(3)\), \(\tilde{\varepsilon} \sim\cN(0, 1)\), and \(X\in\mathbb{R}^d\) is distributed as \(\cN(0, \mathrm{I}_{d})\). 
In this case, \(\theta^* = (1, 1, 0, \ldots, 0)^\T\).
For multidimensional parameters, we compare methods via testing \(H_0: \theta^* = C(1, 1, 0, \ldots, 0)^\T\) versus \(H_1: \theta^* \neq C(1, 1, 0, \ldots, 0)^\T\); thus \(C = 1\) yields empirical size and $C\neq 1$ yields power.
The benchmarks are the same as in Section~\ref{sec:simu:mean}.
For the basis-expansion EPI, we take third-order polynomials of \((\tY, X)\) as the basis functions, as described in Section~\ref{sec:method:basis},
while for the cross-fitted EPI, we again use XGBoost to learn \(s^*\).
All results are averaged over 10000 replications.

\begin{table}[!h]
    \centering
    \caption{Performance of different methods in the linear regression problem.
    The nominal significance level is \(\alpha = 0.1\).}
    \label{tab:linear}
    \parskip0.1in
    \stackunder{
    \resizebox{0.95\linewidth}{!}{
    \begin{tabular}{l *{12}{c}}
        \toprule
         && \multicolumn{5}{c}{\(C = 1\) (Size)} && \multicolumn{5}{c}{\(C = 1.1\) (Power)}\\
        \cmidrule{3-7} \cmidrule{9-13}
        \(\rho\) && 0.1 & 0.5 & 1 & 2 & 5 && 0.1 & 0.5 & 1 & 2 & 5\\
        \midrule
        EPI (Basis)  &&  0.100 & 0.099 & 0.097 & 0.095 & 0.093 && 0.584 & 0.584 & 0.583 & 0.578 & 0.573 \\
        EPI (CF)  &&  0.079 & 0.089 & 0.096 & 0.101 & 0.107 && 0.546 & 0.567 & 0.569 & 0.542 & 0.522 \\ 
        RePPI  &&  0.120 & 0.118 & 0.121 & 0.117 & 0.121 && 0.537 & 0.532 & 0.519 & 0.509 & 0.520 \\ 
        PDC  &&  0.109 & 0.108 & 0.107 & 0.104 & 0.105 && 0.543 & 0.527 & 0.487 & 0.396 & 0.257 \\ 
        PPI  &&  0.108 & 0.104 & 0.106 & 0.104 & 0.102 && 0.469 & 0.456 & 0.410 & 0.300 & 0.162 \\ 
        \(\cls\) &&  0.106 & 0.106 & 0.106 & 0.106 & 0.106 && 0.201 & 0.201 & 0.201 & 0.201 & 0.201 \\ 
        \bottomrule
    \end{tabular}}
    }{\parbox{6in}{\footnotesize
    EPI (Basis), the proposed basis-expansion estimator; EPI (CF), the proposed cross-fitted estimator; RePPI, the recalibrated prediction-powered inference; PDC, the prediction-decorrelated estimator; PPI, the prediction-powered inference; $\cls$, the supervised estimator.}}
\end{table}

Table~\ref{tab:linear} reports results for \((n, d, m) = (1500, 5, 6000)\), where we vary \(\rho\in\{0.1, 0.5, 1, 2, 5\}\) to obtain predictions of varying accuracy (smaller \(\rho\) corresponds to higher prediction quality). 
We observe that all methods have reasonably good size control. Among prediction-powered methods, the basis-expansion EPI, the cross-fitted EPI, and RePPI are least sensitivity to prediction quality (as $\rho$ varies) and maintain relatively high power. The basis-expansion EPI performs best here because \(s^*\) lies in the span of third-order polynomials, allowing it to attain the semiparametric efficiency bound.
The power of both PDC and PPI decreases significantly as the prediction quality worsens.

\subsection{Over-identified supervised score}

Next, we consider mean inference with an over-identified supervised score \(g_{\theta}\).
The data generation process is \(Y = \theta + X_1 + X_2 + \theta \varepsilon\) and \(\tY = X_1 + X_2\), where \(X_1, X_2 \sim \cN(0, \theta^2)\) and \(\varepsilon\sim\cN(0, 1)\).
In addition to \(\E(Y - \theta) = 0\), we also have the moment condition \(\E(Y^2 - 4\theta^2) = 0\).
For estimators based on the prediction-powered inference formulation in \eqref{eq:ppi}, this additional information is not directly incorporable, so we only use \(g_{\theta}(Y, X) = Y - \theta\) for these methods.
For supervised estimators, we consider both the sample mean and the empirical likelihood estimator \citep{Qin1994EL} based on the two-dimensional score \(g_{\theta}(Y, X) = (Y - \theta, Y^2 - 4\theta^2)^\T\).
Our EPI estimators adopt this two-dimensional score: the cross-fitted EPI learns \(\E(Y\mid X)\) and \(\E(Y^2\mid X)\), while the basis-expansion EPI uses the second-order polynomials basis functions.
Here, we report the overall miscoverage, together with the mean squared error and average confidence-interval length.
All results are averaged over 10000 replications.

\begin{figure}[!h]
    \centering
    \includegraphics[width=\linewidth]{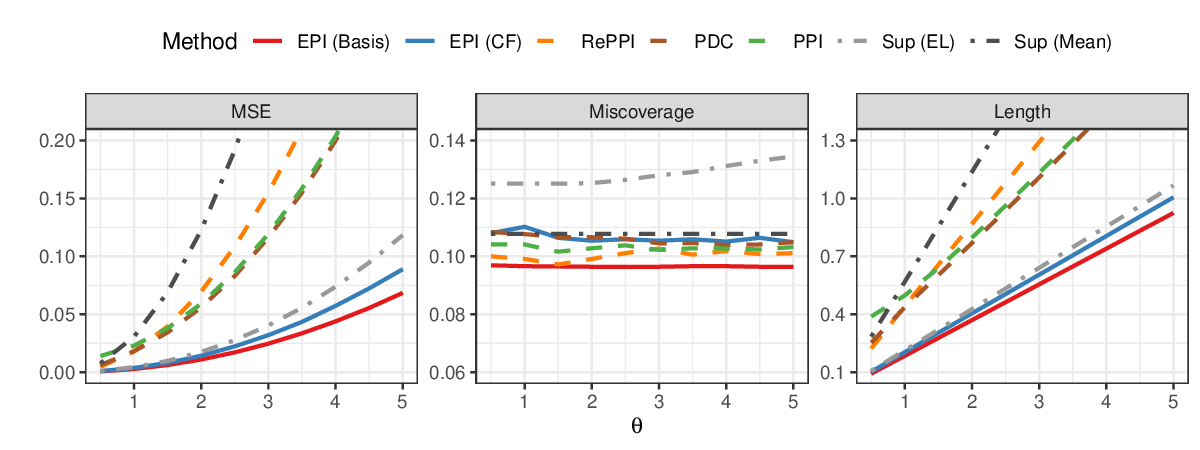}
    \caption{\small Performance of different methods for mean inference under the over-identified \(g_{\theta}\). 
    The nominal miscoverage level is \(\alpha = 0.1\).
    EPI (Basis), the proposed basis-expansion estimator; EPI (CF), the proposed cross-fitted estimator; RePPI, the recalibrated prediction-powered inference; PDC, the prediction-decorrelated estimator; PPI, the prediction-powered inference; $\cls$ (EL), the empirical likelihood-based supervised estimator; $\cls$ (Mean), the sample mean supervised estimator.
    }
    \label{fig:overidentify}
\end{figure}

Figure~\ref{fig:overidentify} presents the results for \((n, m) = (100, 1000)\) and \(\theta\in\{0.5, \ldots, 5\}\), with the nominal miscoverage level \(\alpha = 0.1\). 
The miscoverage probabilities of the different estimators are close to the nominal level, except that the empirical likelihood-based supervised estimator exhibits slightly inflated miscoverage.
In terms of efficiency, estimators that use the over-identified score (the empirical likelihood-based supervised estimator and our EPI estimators) outperform methods that only use \(g_{\theta}(Y, X) = Y - \theta\). 
Furthermore, leveraging predictions, our EPI estimators further improve upon the empirical likelihood-based supervised estimator. 

\subsection{Distribution learning}

Finally, we consider distribution learning for the response. 
The data are generated from \(Y = d_1^{-1/2} \sum_{j = 1}^{d_1} X_j + \sigma\varepsilon\), where \(X = (X_1, \ldots, X_d)^\T\sim \cN(0, I_d)\) and \(\varepsilon\sim \cN(0, 1)\) are independent, and \(d_1 = \lceil 0.15 d\rceil\) is the number of signals. To generate \(\tY\), we train a random-forest predictor on an independent dataset of size $1000$.
The benchmark is the empirical cumulative distribution function \(\hat{F}_{\cls}(y)\).
To improve the overall estimated distribution \(\hat{F}_{\EL}\), we choose a set of points \(\{y_1, \ldots, y_r\}\) that spread across the range of \(\cY\), and set \(s^* = (s^*_{F(y_1)}, \ldots, s^*_{F(y_r)})^\T\) so that \(\hat{F}_{\EL}\) is efficient at \(y_1, \ldots, y_r\).
Accordingly, the basis-expansion and cross-fitting strategies are used to approximate \(s^*\).
We evaluate distribution estimation using the pointwise mean squared error \(\E[\{\hat{F}(y)-F(y)\}^2]\), where \(\hat{F}\) is the estimator produced by each method.
All results are averaged over 10000 replications.

\begin{figure}[!h]
    \centering
    \includegraphics[width=\textwidth]{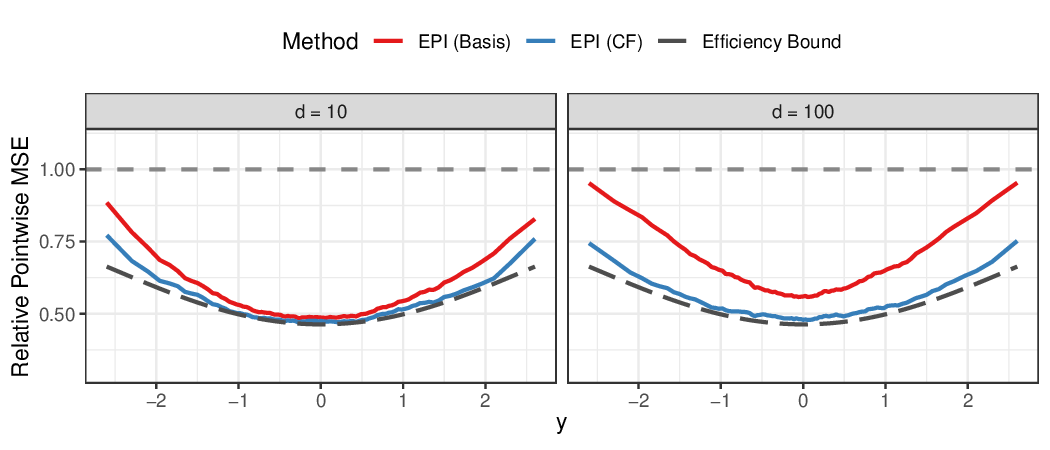}
    \caption{\small Performance of different estimators in the distribution learning problem. Pointwise mean squared errors are displayed as ratios relative to that of \(\hat{F}_{\cls}\).
    The long-dashed curve shows the theoretical relative semiparametric efficiency bound. 
    }
    \label{fig:dist}
\end{figure}

Figure~\ref{fig:dist} presents the results for \((n, m, \sigma) = (1000, 10000, 0.5)\) and \(d \in \{10, 100\}\), where each curve shows the ratio of the pointwise mean squared errors of \(\hat{F}\) to that of \(\hat{F}_{\cls}\). 
A value below $1$ indicates that \(\hat{F}\) is more efficient than \(\hat{F}_{\cls}\).
We observe that our EPI distribution estimators outperform \(\hat{F}_{\cls}\), and the cross-fitted EPI nearly reaches the semiparametric efficiency bound. 

\subsection{Real data analysis}

In this section, we compare different methods on the wine quality data introduced in \citet{CORTEZ2009wine}, including the red and white variants of the Portuguese ``Vinho Verde'' wine. 
For each wine sample, eleven physicochemical properties are recorded, such as sulphates and volatile acidity, and the response is a quality score from 0 to 10 assigned by human experts. 
We apply XGBoost to the red wine dataset to learn the relationship between physicochemical properties and the quality score. 
We then use the learned model to predict the quality score for the white wine dataset, and treat the predicted value as \(\tY\). 
The target of inference is the coefficient vector in the linear regression of \(Y\) on sulphates and volatile acidity, i.e., \(\theta^* = \{\E(X^\T X)\}^{-1}\E(X^\T Y)\) with \(X = (X_\text{sulphates}, X_\text{volatile acidity})^\T\).
We randomly split the \(N = 4898\) samples in the white wine dataset into labeled and unlabeled subsets, where the labeled sample size varies from \(0.1N\) to \(0.5N\), and use the estimate from the full white wine dataset as a proxy for \(\theta^*\).
For this two-dimensional parameter, we compare the benchmarks in Section~\ref{sec:simu:lm} with the proposed basis-expansion EPI (with \(h\) given by second-order polynomials) in terms of mean squared error and coverage.

\begin{table}[!h]
    \centering
    \caption{Performance of different methods on the wine quality data. 
    The nominal coverage level is \(1-\alpha = 0.9\).
    Mean squared errors are reported as ratios relative to \(\hat{\theta}_{\cls}\).}
    \label{tab:wine}
    \parskip0.1in
    \stackunder{
    \resizebox{0.95\linewidth}{!}{
    \begin{tabular}{l *{12}{c}}
        \toprule
        && \multicolumn{5}{c}{Relative MSE} && \multicolumn{5}{c}{Coverage}\\
        \cmidrule{3-7} \cmidrule{9-13}
        $n/N$ && EPI & RePPI & PDC & PPI & $\cls$ && EPI & RePPI & PDC & PPI & $\cls$\\
        \midrule
        0.1 &&  0.21 & 0.28 & 0.25 & 0.27 & 1.00  &&  0.95 & 0.94 & 0.96 & 0.96 & 0.91 \\ 0.2 &&  0.21 & 0.27 & 0.26 & 0.35 & 1.00  &&  0.99 & 0.98 & 0.99 & 0.98 & 0.94 \\ 0.3 &&  0.21 & 0.26 & 0.26 & 0.47 & 1.00  &&  1.00 & 1.00 & 1.00 & 0.99 & 0.96 \\ 0.4 &&  0.21 & 0.24 & 0.26 & 0.67 & 1.00  &&  1.00 & 1.00 & 1.00 & 0.99 & 0.98 \\ 0.5 &&  0.19 & 0.24 & 0.25 & 1.16 & 1.00  &&  1.00 & 1.00 & 1.00 & 0.98 & 0.99 \\
        \bottomrule
    \end{tabular}}}{
    \parbox{6in}{\footnotesize 
    EPI, the proposed basis-expansion estimator; RePPI, the recalibrated prediction-powered inference; PDC, the prediction-decorrelated estimator; PPI, the prediction-powered inference; $\cls$, the supervised estimator.}}
\end{table}

\begin{figure}[!h]
    \centering
    \includegraphics[width=\linewidth]{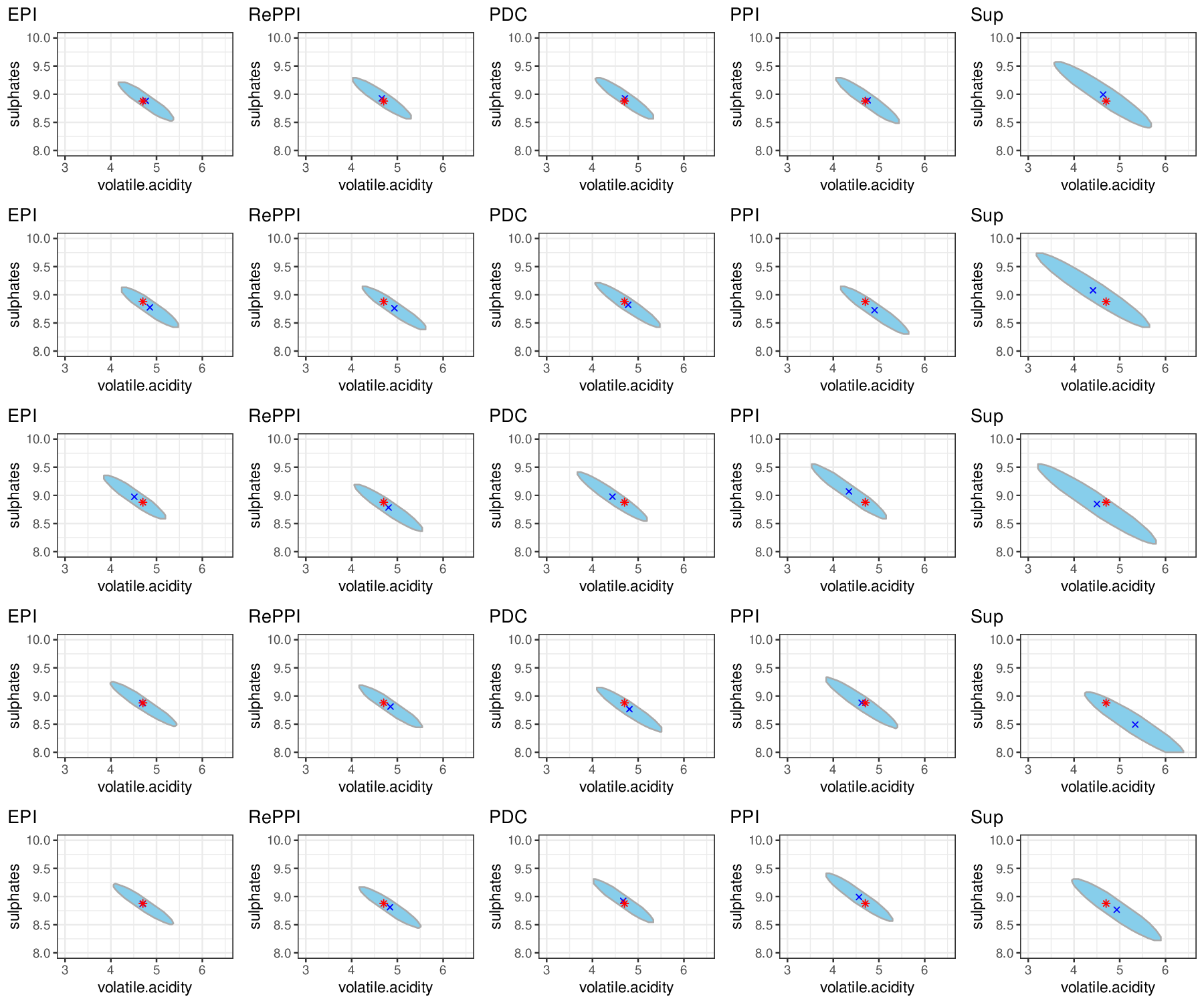}
    \caption{\small Confidence regions of different methods on the wine quality data when \(n = 0.2N\), with nominal coverage \(1-\alpha = 0.9\).
    Each row corresponds to one replication.
    The shaded area is the confidence region, and the red star and blue cross denote the proxy \(\theta^*\) (full-sample estimate) and the replication-specific point estimate, respectively. 
    EPI, the proposed basis-expansion estimator; RePPI, the recalibrated prediction-powered inference; PDC, the prediction-decorrelated estimator; PPI, the prediction-powered inference; $\cls$, the supervised estimator.}
    \label{fig:wine}
\end{figure}

Table~\ref{tab:wine} summarizes results for the wine quality data. 
The reported coverage becomes more conservative as the labeled proportion increases; this is partly an artifact of using the full-sample estimate as a proxy for \(\theta^*\).
In terms of efficiency, EPI consistently achieves a lower mean squared error than the competing methods. 
Figure~\ref{fig:wine} shows five replications of the confidence regions when \(n = 0.2N\); the region sizes align with the relative mean squared error results in Table~\ref{tab:wine}. 

\section{Concluding remarks}\label{sec:con}

We proposed an empirical likelihood framework for prediction-powered inference.
Our method, EPI, combines a supervised estimating equation based on labeled outcomes with auxiliary moment conditions built from predictions, and uses empirical likelihood to choose weights on the labeled sample.
These weights reconcile supervised and prediction-based information while keeping the target defined by the original estimating equation.
Under mild regularity conditions, any valid auxiliary construction yields an estimator whose asymptotic variance is no larger than that of the supervised estimator, and the semiparametric efficiency bound is attained when the centered auxiliary function spans the predictable component \(s^*(\tY,X)=\E\{g_{\theta^*}(Y,X)\mid \tY,X\}\).
The same empirical likelihood construction also induces a calibrated empirical distribution, which allows distributional functionals to be treated within the same prediction-powered framework.

Several directions remain open.
First, throughout we take the prediction model as externally provided.
A fully semi-supervised version of EPI, where the predictor is learned from the same data and interacts with the empirical likelihood optimization, would connect more directly to existing semi-supervised estimation and raises questions about joint training and cross-fitting.
Second, our analysis focused on the canonical setting in which labels are missing completely at random.
Recent work in semi-supervised and prediction-powered inference allows label probabilities to depend on covariates, predictions, or other proxies, and considers covariate or label shift \citep[e.g.,][]{Zhang2023decay,Testa2025decay}.
Adapting EPI to such mechanisms, for example using inverse-probability weighting or augmented empirical likelihood constraints, is an important topic for future research.

\putbib
\end{bibunit}

\begin{bibunit}
    
\newpage
\begin{center}
  {\Large\bf Supplementary Material for ``Empirical Likelihood Meets Prediction-Powered Inference''}  
\end{center}

\def\thesection{S.\arabic{section}}
\def\thelem{S.\arabic{lem}}
\def\theprop{S.\arabic{prop}}
\def\thethm{S.\arabic{thm}}
\def\theequation{S.\arabic{equation}}
\def\thetable{S\arabic{table}}
\def\thefigure{S\arabic{figure}}
\def\thealgocf{S\arabic{algocf}}
\setcounter{section}{0}
\setcounter{equation}{0}
\setcounter{thm}{0}
\setcounter{lem}{0}
\setcounter{prop}{0}
\setcounter{figure}{0}
\setcounter{table}{0}

This supplementary material includes general theoretical results for the proposed empirical likelihood-based prediction-powered inference (EPI) method (Section~\ref{supp:theory}), implementation details (Section~\ref{supp:simu}), additional numerical results (Section~\ref{supp:real}), and all technical proofs (Section~\ref{supp:proof}).

\section{General theory of EPI}\label{supp:theory}

This section presents general theoretical results for EPI, covering both just-identified and over-identified supervised scores and all constructions considered in the paper (the ideal case, the centered case, and the cross-fitted case).

Recall that the profile empirical likelihood with supervised score \(g_{\theta}\) (of dimension \(q \geq p\)) is
\begin{equation*}
    \begin{split}
        \ell_n(\theta):=\max\Big\{\prod_{i=1}^n w_i: ~&w_i\geq 0, \sum_{i=1}^n w_i=1,\sum_{i=1}^n w_i g_{\theta}(Y_i,X_i) = 0,\\
        &\sum_{k=1}^K\sum_{i \in\cI_k} w_i \big\{\hat{h}^{(-k)}(\tY_i, X_i) - \frac{1}{n_K+m_K}\sum_{\ell \in \cI_k\cup\cJ_k} \hat{h}^{(-k)}(\tY_\ell, X_\ell)\big\} = 0\Big\},
    \end{split}
\end{equation*}
which corresponds to the cross-fitted case. 
In particular, if \(\hat{h}^{(-k)}(\tY, X) = h(\tY, X)\) for a given \(h\) and all \(k = 1, \ldots, K\), then we recover the centered construction; moreover, when \(m = \infty\), this formulation reduces to the ideal case.
Maximizing \(\ell_n(\theta)\) yields the estimator \(\hat{\theta}_{\EL}\). We next state general results on asymptotic normality, the limiting distribution of the empirical likelihood ratio, and weak convergence of the calibrated distribution function.

\begin{thm}\label{THM:SUPP:NORM}
    Suppose that Assumptions~\ref{asmp:EL}--\ref{asmp:learn-h} hold. Additionally, assume that \(\E\{\|\h\|^3\} < \infty\), and that \(\cov((g_{\theta^*}^\T, \h^\T)^\T)\) is positive definite. 
    If $n/(n+m)\to \gamma$ for some constant $\gamma\in[0,1]$, then
    \begin{align*}
        \sqrt{n}(\hat{\theta}_{\EL}-\theta^*) \rightsquigarrow \mathcal{N}(0,\Sigma_{\h}),
    \end{align*}
    where
        \begin{equation*}
        \Sigma_{h} = \left(J^{\T}U_h^{-1}J \left[J^{\T}U_h^{-1}J + \gamma J^{\T}U_h^{-1} \E(g_{\theta^*} h^\T) \{\cov(h)\}^{-1} \E(hg_{\theta^*}^\T) U_h^{-1}J\right]^{-1} J^{\T}U_h^{-1}J\right)^{-1},
    \end{equation*}
    with \(J=\E(\partial g_{\theta^*}/\partial\theta)\) and \(U_h = \cov(g_{\theta^*}) - \E(g_{\theta^*}h^\T) \{\cov(h)\}^{-1} \E(hg_{\theta^*}^\T)\). 
    Furthermore, \(\Sigma_\h \preceq [J^\T\{\cov(g_{\theta^*})\}^{-1} J]^{-1}\) holds when 
    \[
        \gamma \leq (1+\lambda_{\max}[(\cov{\h})^{1/2} P\h g_{\theta^*}^{\top} \{U_\h^{-1}-U_\h^{-1}J(J^\T U_\h^{-1}J)^{-1}J^\T U_\h^{-1}\} Pg_{\theta^*}\h^\top (\cov{\h})^{1/2}])^{-1}.
    \]
\end{thm}

\begin{thm}\label{THM:SUPP:LR}
    Under the conditions of Theorem~\ref{THM:SUPP:NORM},   
    \[
    T_n(\theta^*) := -2\log\{\ell_n(\theta^*)/\ell_n(\hat{\theta}_{\EL})\} \rightsquigarrow \sum_{j = 1}^p \lambda_j \xi_j^2,
    \] 
    where \(\xi_1, \ldots, \xi_p\) are independent standard normal random variables, and \(\lambda_1, \ldots, \lambda_p\) are eigenvalues of 
    \[
        I+ \gamma (J^\top U_\h^{-1}J)^{-1/2}J^\top U_\h^{-1}\E(g_{\theta^*}\h^\T)\{\cov(\h)\}^{-1} \E(\h g_{\theta^*}^{\T}) U_\h^{-1}J(J^\T U_\h^{-1}J)^{-1/2}.
    \]
\end{thm}

\begin{thm}\label{THM:SUPP:DIST}
    Under the conditions of Theorem~\ref{THM:SUPP:NORM}, for any \((y, x) \in \cY\times \cX\), 
    \[
    \sqrt{n}\left\{\hat{F}_{\EL}(y, x) - F(y, x)\right\} \rightsquigarrow \mathcal{N}\left(0, \sigma^2_{\h}(y, x)\right),
    \]
    where 
    \[
        \sigma^2_{h}(y, x) = F(y,x)\{1-F(y,x)\}-\rho_{h I(y, x)}^\T \{\cov(h)\}^{-1}\rho_{h I(y, x)} -a_h^\T\Pi_{U; h} a_h + \gamma b_h^\T\{\cov(h)\}^{-1}b_h,
    \]
    with $\rho_{gI(y, x)} = \E \{g_{\theta^*}(Y,X) I(Y\leq y, X\leq x)\}$, $\rho_{h I(y, x)}=\cov\{h(\tY, X), I(Y\leq y, X\leq x)\}$, $\Pi_{U;h}=U_h^{-1}-U_h^{-1}J(J^\T U_h^{-1}J)^{-1}J^\T U_h^{-1}$, $a_h = \rho_{gI(y, x)} - \E(g_{\theta^*}h^{\T})\{\cov(h)\}^{-1}\rho_{h I(y, x)}$ and $b_h = \E(h g^\T_{\theta^*})\Pi_{U; h} a-\rho_{h I(y, x)}$.
    Furthermore, \(\sigma^2_{\h}(y, x) \leq F(y,x)\{1-F(y,x)\}\) holds when 
    \[
    \gamma b_\h^\top(\cov{\h})^{-1}b_\h\leq \rho_{\h I(y, x)}^\top (\cov{\h})^{-1}\rho_{\h I(y, x)}+a_\h^\top\Pi_{U;h} a_\h.
    \]
\end{thm}

For these results, note that (i) when \(\gamma = 0\), Theorems~\ref{THM:SUPP:NORM}--\ref{THM:SUPP:DIST} align with the classical results of \cite{Qin1994EL}, and (ii) for a just-identified supervised score \(g_{\theta}\) (i.e., \(q = p\), so that \(J\) is invertible), Theorems~\ref{THM:SUPP:NORM}--\ref{THM:SUPP:DIST} recover Theorems~\ref{thm:ss}, \ref{thm:LR-ss}, and \ref{thm:dist}, respectively.

\section{Implementations}\label{supp:simu}

\subsection{Inference based on the empirical likelihood ratio}

Theorems~\ref{thm:LR-ss}, \ref{thm:crossfit}, and \ref{THM:SUPP:LR} show that the empirical likelihood ratio statistic converges to a weighted sum of independent chi-squared random variables.
We use the centered construction with a just-identified supervised score \(g_{\theta}\) as an illustrative example; the other cases can be handled analogously.

We estimate the eigenvalues \(\lambda_1, \ldots, \lambda_p\) of \(\Lambda_h\), where 
\[
        \Lambda_h = I_p + \gamma (J^\T U_h^{-1}J)^{1/2} J^{-1} \E(g_{\theta^*} h^\T) \{\cov(h)\}^{-1} \E(h g_{\theta^*}^\T) J^{-\T} (J^\T U_h^{-1} J)^{1/2},
\]
and \(U_h = \cov(g_{\theta^*}) - \E(g_{\theta^*}h^\T) \{\cov(h)\}^{-1} \E(hg_{\theta^*}^\T)\).
Natural estimators of \(\lambda_1, \ldots, \lambda_p\) are given by the eigenvalues of the plug-in estimator \(\hat{\Lambda}_h\) of $\Lambda_h$:
\[
\hat{\Lambda}_h = I_p + \gamma_n (\hat{J}^\T \hat{U}_h^{-1}\hat{J})^{1/2} \hat{J}^{-1} \hat{\cov}(g_{\theta^*}, h) \{\hat{\cov}(h)\}^{-1} \hat{\cov}(h, g_{\theta^*}) \hat{J}^{-\T} (\hat{J}^\T \hat{U}_h^{-1} \hat{J})^{1/2},
\]
where \(\gamma_n = n/(n+m)\), \(\hat{U}_h = \hat{\cov}(g_{\theta^*}) - \hat{\cov}(g_{\theta^*}, h) \{\hat{\cov}(h)\}^{-1} \hat{\cov}(h, g_{\theta^*})\), 
\[\begin{split}
  &\hat{J} = \frac{1}{n}\sum_{i = 1}^n \partial g_{\hat{\theta}}(Y_i, X_i)/\partial \theta,\\
  &\hat{\cov}(g_{\theta^*}) = \frac{1}{n}\sum_{i = 1}^n g_{\hat{\theta}}(Y_i, X_i)g_{\hat{\theta}}^\T(Y_i, X_i) - \left\{\frac{1}{n}\sum_{i = 1}^n g_{\hat{\theta}}(Y_i, X_i)\right\}\left\{\frac{1}{n}\sum_{i = 1}^n g_{\hat{\theta}}(Y_i, X_i)\right\}^\T,\\
  &\hat{\cov}(h) = \frac{1}{n}\sum_{i = 1}^n {h}(\tY_i, X_i){h}^\T(\tY_i, X_i) - \left\{\frac{1}{n}\sum_{i = 1}^n {h}(\tY_i, X_i)\right\}\left\{\frac{1}{n}\sum_{i = 1}^n {h}(\tY_i, X_i)\right\}^\T,\\
  &\hat{\cov}(g_{\theta^*}, h) = \frac{1}{n}\sum_{i = 1}^n g_{\hat{\theta}}(Y_i, X_i){h}^\T(\tY_i, X_i) - \left\{\frac{1}{n}\sum_{i = 1}^n g_{\hat{\theta}}(Y_i, X_i)\right\}\left\{\frac{1}{n}\sum_{i = 1}^n {h}(\tY_i, X_i)\right\}^\T,
\end{split}\]
and \(\hat{\theta}\) is any \(\sqrt{n}\)-consistent estimator of \(\theta^*\), (e.g., the EPI estimator). 
We then compute the eigenvalues of \(\hat{\Lambda}_h\) via singular value decomposition to obtain \(\hat{\lambda}_1, \ldots, \hat{\lambda}_p\). Substituting these eigenvalues into the limiting distribution yields critical values and the resulting confidence intervals.

\subsection{Computations}

We compute the EPI estimator \(\hat{\theta}_{\EL; h}\) and the empirical likelihood ratio statistic \(T_n(\theta) = -2\log\{\ell_n(\theta^*; h)/\ell_n(\hat{\theta}_{\EL; h}; h)\}\) through Lagrange multipliers, following standard empirical likelihood arguments \citep{Owen2001EL}.
The cross-fitted EPI estimator can be obtained similarly.
Let the dimensions of \(g_{\theta}\) and \(h\) be \(q\) and \(r\), respectively, with \(q\geq p\).
Denote \(h^c(\tY, X) := h(\tY, X) - (n+m)^{-1}\sum_{j = 1}^{n+m} h(\tY_j, X_j)\).
Consider the Lagrangian 
\[\begin{split}
    &\cL(w_1, \ldots, w_n, \theta, \lambda, t_1, t_2; h) = \\ 
    &\qquad  -\sum_{i = 1}^n \log(w_i) + \lambda\left(\sum_{i = 1}^n w_i - 1\right) + n t_1^\T \sum_{i = 1}^n w_i g_\theta(Y_i, X_i) + n t_2^\T \sum_{i = 1}^n w_i h^c(\tY_i, X_i),
\end{split}\]
where \(t_1\in\bR^{q}\), \(t_2\in\bR^r\), and \(\lambda\) are Lagrange multipliers.
Differentiating with respect to \(w_i\) gives
\[\begin{split}
    &\frac{\partial \cL}{\partial w_i} = -\frac{1}{w_i} + \lambda + n t_1^\T g_\theta(Y_i, X_i) + n t_2^\T h^c(\tY_i, X_i) = 0, \quad \sum_{i = 1}^n w_i \frac{\partial \cL}{\partial w_i} = - n + \lambda = 0,\\
\end{split}\]
and hence
\[
\lambda = n,\quad w_i = \frac{1}{n\{1 + t_1^\T g_\theta(Y_i, X_i) + t_2^\T h^c(\tY_i, X_i)\}},
\]
subject to the constraints
\[
\frac{1}{n}\sum_{i = 1}^n \frac{g_\theta(Y_i, X_i)}{1 + t_1^\T g_\theta(Y_i, X_i) + t_2^\T h^c(\tY_i, X_i)} = 0,\quad \frac{1}{n}\sum_{i = 1}^n \frac{h^c(\tY_i, X_i)}{1 + t_1^\T g_\theta(Y_i, X_i) + t_2^\T h^c(\tY_i, X_i)} = 0,
\]
which determine \(t_1\) and \(t_2\) as functions of \(\theta\); see \citet{Qin1994EL} for related arguments.
To calculate \(t_1\) and \(t_2\) for a given \(\theta\), we use the Newton method described in Chapter~3.14 of \citet{Owen2001EL}.
The negative empirical log-likelihood can then be written as 
\[
\log \ell_n(\theta; h) = -\sum_{i = 1}^n \log\{1 + t_1^\T(\theta) g(Y_i, X_i; \theta) + t_2^\T(\theta) h^c(\tY_i, X_i)\} - n\log n.
\]
We maximize \(\log \ell_n(\theta; h)\) to obtain \(\hat{\theta}_{\EL;h}\), and compute the empirical likelihood ratio statistic \(T_n(\theta) = 2\log\{\ell_n(\hat{\theta}_{\EL; h}; h)\}-2\log\{\ell_n(\theta; h)\}\).

\section{Additional real data analysis}\label{supp:real}

We analyze the homeless-count data from Los Angeles County \citep{Kriegler2010homeless}, collected by the Los Angeles Homeless Services Authority in 2004--2005. 
A stratified spatial sampling was conducted over 2054 census tracts. Specifically, 244 tracts (``hot tracts''), believed to have a high concentration of homeless individuals, were preselected and visited. In addition, $n=265$ of the remaining tracts were randomly selected and visited, leaving $m=1545$ tracts unvisited. 
The response \(Y\) is the number of homeless individuals per tract. The covariates \(X\) include demographic and socioeconomic characteristics; we use seven tract-level predictors available for all tracts: \texttt{Industrial}, \texttt{Residential}, \texttt{PctVacant}, \texttt{Commercial}, \texttt{PctOwnerOcc}, \texttt{PctMinority}, and \texttt{MedianHouseholdIncome}.
We train an XGBoost model \(f\) using data from the 244 hot tracts and use it to produce predictions \(\tY=f(X)\) for all non-hot tracts.

\begin{table}[!h]
    \centering
    \caption{Quantile inference results for the homeless data analysis.}
    \label{tab:homeless}
    \parskip0.1in
    \stackunder{
    \begin{tabular}{l *{9}{c}}
        \toprule
        && \multicolumn{2}{c}{\(\tau = 0.3\)} && \multicolumn{2}{c}{\(\tau = 0.5\)} && \multicolumn{2}{c}{\(\tau = 0.7\)}\\
        \cmidrule{3-4} \cmidrule{6-7} \cmidrule{9-10}
        Method && \(\hat{\theta}\) & \(95\%\)-CI && \(\hat{\theta}\) & \(95\%\)-CI && \(\hat{\theta}\) & \(95\%\)-CI\\
        \midrule
        EPI &&  5.0 & $[ 4.0,  7.0]$ && 12.0 & $[10.0, 15.0]$ && 25.0 & $[20.0, 27.0]$ \\ 
        RePPI &&  4.7 & $[ 3.4,  6.0]$ && 12.3 & $[\ph 9.9, 14.8]$ && 23.6 & $[19.9, 27.4]$ \\ 
        PDC &&  5.0 & $[ 3.8,  6.2]$ && 12.0 & $[\ph 9.4, 14.6]$ && 24.0 & $[20.2, 27.8]$ \\ 
        PPI &&  6.3 & $[ 6.2,  6.3]$ && 15.0 & $[14.9, 15.1]$ && 23.7 & $[23.6, 23.8]$ \\
        $\cls$ &&  5.0 & $[ 3.8,  6.2]$ && 12.0 & $[\ph 9.4, 14.6]$ && 24.0 & $[20.2, 27.8]$ \\ 
        \bottomrule
    \end{tabular}}{
    \parbox{6in}{\footnotesize
    EPI, the proposed basis-expansion estimator; RePPI, the recalibrated prediction-powered inference; PDC, the prediction-decorrelated estimator; PPI, the prediction-powered inference; $\cls$, the supervised estimator.}}
\end{table}

\begin{figure}[!h]
    \centering
    \includegraphics[width=0.8\linewidth]{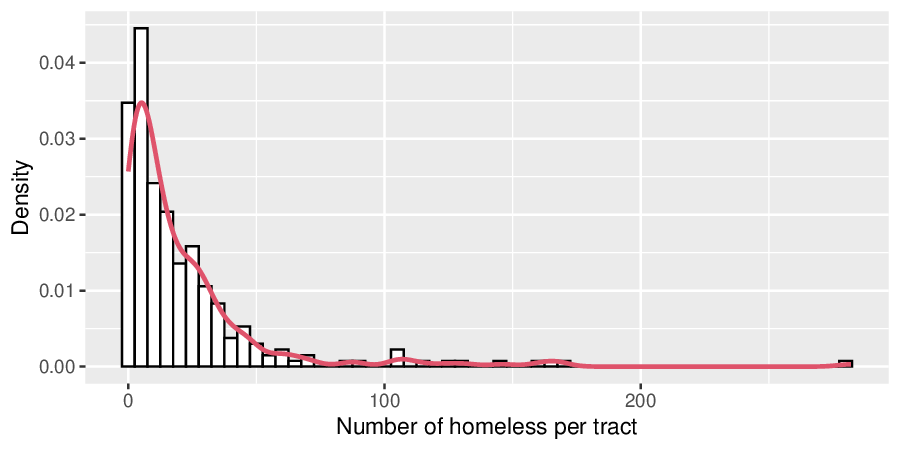}
    \caption{\small Histogram of homeless counts per tract among the visited non-hot tracts in the homeless data analysis.}
    \label{fig:homeless}
\end{figure}

We conduct inference on the \(\tau\)th quantile of the homeless count per tract among non-hot tracts. 
The EPI quantile estimator is obtained by inverting the estimated distribution function \(\hat{F}_{\EL}(y)\).
Table~\ref{tab:homeless} reports results for \(\tau\in\{0.3, 0.5, 0.7\}\) with nominal confidence level \(0.95\). 
The supervised estimator and the prediction-powered methods yield confidence intervals (CIs) that are approximately symmetric around the point estimate, which may be poorly suited to the highly skewed response distribution (Figure~\ref{fig:homeless}). 
In contrast, EPI produces asymmetric confidence intervals that better reflect the skewness.

\section{Proofs}\label{supp:proof}

We first introduce some notation.
For a sample $\{Z_i\}_{i=1}^n$ of size $n$ and a (possibly data-dependent) function $f$, let $P_nf=n^{-1}\sum_{i=1}^nf(Z_i)$. For example, $P_{N_k} \hat{h}^{(-k)}=N_k^{-1}\{\sum_{i\in I_k}\hat{h}^{(-k)}(\tY_i,X_i)+\sum_{j\in J_k}\hat{h}^{(-k)}(\tY_j,X_j)\}$ with $N_k=n_k+m_k$, $n_k=|I_k|$ and $m_k=|J_k|$. Here $\hat{h}^{(-k)}$ is data-dependent and the sample is $\{Z_i=(\tY_i,X_i):i\in I_k\cup J_k\}$. When the function depends on the sample index $i$, say $f_i$, with a slight abuse of notation, let $P_nf_i=n^{-1}\sum_{i=1}^nf_i(Z_i)$. 
For $Z\sim P$ and a function $f$, let $Pf=E_P\{f(Z)\}$.

\subsection*{Asymptotic normality of EPI estimators (Theorem~\ref{THM:SUPP:NORM})}

We maximize $\prod_{i=1}^n w_i$, or equivalently $\sum_{i=1}^n \log w_i$, subject to restrictions
\[
w_i\geq 0,\ \sum_{i=1}^n w_i=1,\ \sum_{i=1}^n w_i g_\theta(Y_i,X_i)=0,\text{ and }\sum_{i=1}^n w_i \hat{h}^c_i(\tilde{Y}_i, X_i)=0,
\]
where
\[
\hat{h}^c_i(\tilde{Y}_i, X_i) = \sum_{k=1}^K\{\hat{h}^{(-k)}(\tilde{Y}_i, X_i)-P_{N_k} \hat{h}^{(-k)}\} I\left\{i \in I_k\right\}.
\]
Consider the Lagrangian
\[
\sum_{i=1}^n \log w_i + \lambda(1-\sum_{i=1}^n w_i) -nt_g^\top\sum_{i=1}^n w_i g_\theta(Y_i,X_i) -nt_h^\top\sum_{i=1}^n w_i\hat{h}^c_i(\tilde{Y}_i, X_i),
\]
where $\lambda\in\mathbb{R}$, $t_g\in\mathbb{R}^{q}$ and $t_h\in\mathbb{R}^{r}$ are Lagrange multipliers. Differentiating with respect to $w_i$ yields $\lambda=n$ and
\[
w_i=\frac{1}{n\{1+t_g^\top g_\theta(Y_i,X_i)+t_h^\top\hat{h}^c_i(\tilde{Y}_i, X_i)\}},
\]
under the restriction $\phi_{n}(\theta;t_g,t_h)=0$, where
\begin{align*}
\phi_{n}(\theta;t_g,t_h)=\begin{pmatrix}
P_n\frac{g_\theta}{1+t_g^\top g_\theta+t_h^\top\hat{h}^c_i}\\
P_n\frac{\hat{h}^c_i}{1+t_g^\top g_\theta+t_h^\top\hat{h}^c_i}
\end{pmatrix}.
\end{align*}
The profile empirical log-likelihood is therefore
\[
\ell_n(\theta)=-\sum_{i=1}^n\log\{1 + t_g^\top g_\theta(Y_i,X_i) + t_h^\top \hat{h}^c_i(\tilde{Y}_i, X_i)\}-n\log n.
\]
Using arguments similar to \cite{Qin1994EL}, $\hat{\theta}=\arg\max\ell_n(\theta)$, together with $\hat{t}_g=t_g(\hat{\theta})$ and $\hat{t}_h=t_h(\hat{\theta})$, satisfies both $\phi_{n}(\theta;t_g,t_h)=0$ and $\varphi_{n}(\theta;t_g,t_h)=0$, where
\[
\varphi_{n}(\theta;t_g,t_h)=P_n\frac{\frac{\partial}{\partial\theta} g_{\theta}^\top t_g}{1+t_g^\top g_\theta+t_h^\top\hat{h}^c_i}.
\]

Applying a Taylor expansion yields
\begin{align*}
\begin{pmatrix}
\sqrt{n} \hat{t}_g \\ 
\sqrt{n} \hat{t}_h \\ 
\sqrt{n}(\hat{\theta}-\theta^*)
\end{pmatrix}
=-\begin{pmatrix}
- A_n & B_n \\
B_n^\top & 0
\end{pmatrix}^{-1}
\begin{pmatrix}
c_n \\
0
\end{pmatrix} + o_P(1),
\end{align*}
where
\[
A_n=\begin{pmatrix}
P_n g_{\theta^*}g_{\theta^*}^\top & P_n g_{\theta^*}\hat{h}^{c\top}_i \\
P_n \hat{h}^{c}_ig_{\theta^*}^\top & P_n \hat{h}^c_i\hat{h}^{c\top}_i
\end{pmatrix},
\text{ }
B_n=\begin{pmatrix}
P_n\frac{\partial}{\partial\theta^\top}g_{\theta^*} \\
0
\end{pmatrix}
\text{ and }
c_n=\begin{pmatrix}
\sqrt{n}P_n g_{\theta^*} \\
\sqrt{n}P_n \hat{h}^c_i
\end{pmatrix}.
\]

By Assumption \ref{asmp:learn-h}, for each $k$, conditional on the training data used to construct $\hat{h}^{(-k)}$, we have $P\|\hat{h}^{(-k)} - \h\|^2$ converges in probability to $0$.

\noindent\textbf{Fact I:} $c_n \rightsquigarrow\mathcal{N}(0,\Sigma)$, where
\[
\Sigma
=\begin{pmatrix}
P g_{\theta^*}g_{\theta^*}^\top & (1-\gamma)P g_{\theta^*}\h^\top \\
(1-\gamma)P \h g_{\theta^*}^\top & (1-\gamma)\cov(\h)
\end{pmatrix}.
\]

\noindent\textit{Proof of Fact I.} We decompose
\begin{align*}
\sqrt{n}P_n \hat{h}^c_i
& =\sqrt{n}n^{-1} \sum_{k\in[K]}\sum_{i\in I_k}\{\hat{h}^{(-k)}(\tilde{Y}_i, X_i)-P_{N_k} \hat{h}^{(-k)}\} \\
& =\sqrt{n}(P_n\h-P\h) - \sqrt{n}n^{-1}\sum_{k\in[K]}n_k(P_{N_k}\h-P\h) \\
& ~~~~~~~~~~+ \sqrt{n}n^{-1}\sum_{k\in[K]}n_k\{P_{n_k}(\hat{h}^{(-k)}-\h)-P(\hat{h}^{(-k)}-\h)\} \\
& ~~~~~~~~~~- \sqrt{n}n^{-1}\sum_{k\in[K]}n_k\{P_{N_k}(\hat{h}^{(-k)}-\h)-P(\hat{h}^{(-k)}-\h)\}.
\end{align*}
By Lemma 19.24 of \cite{vanderVaart-1998-pxvi+443} (see also the proof of Theorem 3 in \cite{Ji2025RePPI}),
\[
\sqrt{n}\{P_{n_k}(\hat{h}^{(-k)}-\h)-P(\hat{h}^{(-k)}-\h)\}=o_P(1)\text{ and }\sqrt{N_k}\{P_{N_k}(\hat{h}^{(-k)}-\h)-P(\hat{h}^{(-k)}-\h)\}=o_P(1).
\]
Hence,
\begin{align*}
\sqrt{n}P_n \hat{h}^c_i &= \sqrt{n}(P_n\h-P\h) - \sqrt{n}n^{-1}\sum_{k\in[K]}n_k(P_{N_k}\h-P\h) + o_P(1).
\end{align*}
Notice that
\[
\sqrt{n}P_n \hat{h}^c_i = \sum_{i=1}^n \sqrt{n}(n^{-1}-N^{-1})\{\h(\tY_i,X_i)-P\h\}-\sqrt{n}N^{-1}\sum_{i=n+1}^{n+m}\{\h(\tY_i,X_i)-P\h\}+o_P(1)
\]
and
\[
\sqrt{n}P_n g_{\theta^*}=\sqrt{n}n^{-1}\sum_{i=1}^n g_{\theta^*}(Y_i,X_i).
\]
Assuming $n/N\to \gamma$, Fact I follows by the Lindeberg central limit theorem. \qed

\noindent\textbf{Fact II:}
\[
A_n\xrightarrow{P}A\equiv\begin{pmatrix}
P g_{\theta^*}g_{\theta^*}^\top & P g_{\theta^*}\h^\top \\
P \h g_{\theta^*}^\top & \cov(\h)
\end{pmatrix}
\text{ and }
B_n\xrightarrow{P}B\equiv\begin{pmatrix}
P\frac{\partial}{\partial\theta^\top}g_{\theta^*} \\
0
\end{pmatrix}.
\]

\noindent\textit{Proof of Fact II.} The limits $P g_{\theta^*}g_{\theta^*}^\top$ and $P\frac{\partial}{\partial\theta^\top}g_{\theta^*}$ follow from the law of large numbers. For $P g_{\theta^*}\h^\top$ and $\cov(\h)$, arguments analogous to those used in the proof of Fact I can be used, and the proof is omitted. \qed

Combining Facts I--II, standard algebra yields
\begin{align*}
\sqrt{n}(\hat{\theta}-\theta^*)
& =-(B^\top A^{-1}B)^{-1}B^\top A^{-1}c_n \\
&\rightsquigarrow\mathcal{N}(0,\Omega^{-1}),
\end{align*}
where, with $J=P \frac{\partial}{\partial \theta^\top} g_{\theta^*}$ and $U_\h=Pg_{\theta^*} g_{\theta^*}^{\top}-P g_{\theta^*}\h^{\top}(\cov{\h})^{-1} P \h g_{\theta^*}^{\top}$,
\begin{align*}
\Omega = J^\top U_\h^{-1}J\Big\{J^\top U_\h^{-1}J+ \gamma J^\top U_\h^{-1}P g_{\theta^*}\h^{\top}(\cov{\h})^{-1} P \h g_{\theta^*}^{\top} U_\h^{-1}J\Big\}^{-1}J^\top U_\h^{-1}J.
\end{align*}
In particular, when $J$ is invertible,
\[
\Omega=P \frac{\partial}{\partial \theta} g_{\theta^*}^\top\{ Pg_{\theta^*} g_{\theta^*}^{\top}-(1-\gamma) P g_{\theta^*}\h^{\top}(\cov{\h})^{-1} P \h g_{\theta^*}^{\top}\}^{-1}P \frac{\partial}{\partial \theta^{\top}} g_{\theta^*}.
\]
Moreover, \(\Omega \succeq J^\top (P g_{\theta^*}g_{\theta^*}^\T)^{-1} J\) when
\[
    \gamma \leq (1+\lambda_{\max}[(\cov{\h})^{1/2} P\h g_{\theta^*}^{\top} \{U_\h^{-1}-U_\h^{-1}J(J^\T U_\h^{-1}J)^{-1}J^\T U_\h^{-1}\} Pg_{\theta^*}\h^\top (\cov{\h})^{1/2}])^{-1}.
\]

\subsection*{Chi-squared-type limit of EPI tests (Theorem~\ref{THM:SUPP:LR})}

The log-empirical likelihood ratio test statistic is
\begin{align*}
2\ell_n(\hat{\theta})-2\ell_n(\theta_0)
&=2\sum_{i=1}^n\log\{1 + \hat{t}_{0g}^\top g_{\theta_0}(Y_i,X_i) + \hat{t}_{0h}^\top \hat{h}^c_i(\tilde{Y}_i, X_i)\}\\
&~~~~~~~~~~-2\sum_{i=1}^n\log\{1 + \hat{t}_g^\top g_{\hat{\theta}}(Y_i,X_i) + \hat{t}_h^\top \hat{h}^c_i(\tilde{Y}_i, X_i)\},
\end{align*}
where $\hat{t}_{0g}$ and $\hat{t}_{0h}$ are the estimated Lagrange multipliers under $H_0:\theta=\theta_0$, analogously to $\hat{t}_{g}$ and $\hat{t}_{h}$.

With a slight abuse of notation, we continue to use $A$, $B$ and $c_n$ from the previous section, with $\theta^*$ replaced by $\theta_0$. By the same arguments,
\[
\begin{pmatrix}
\sqrt{n}\hat{t}_{0g}\\
\sqrt{n}\hat{t}_{0h}
\end{pmatrix}
=A^{-1} c_n+o_P(1).
\]
Moreover, a Taylor expansion gives
\begin{align*}
2\sum_{i=1}^n\log\{1 + \hat{t}_{0g}^\top g_{\theta_0}(Y_i,X_i) + \hat{t}_{0h}^\top \hat{h}^c_i(\tilde{Y}_i, X_i)\} 
&= c_n^\top A^{-1} c_n + o_P(1). 
\end{align*}

Similarly,
\begin{align*}
2\sum_{i=1}^n\log\{1 + \hat{t}_g^\top g_{\hat{\theta}}(Y_i,X_i) + \hat{t}_h^\top \hat{h}^c_i(\tilde{Y}_i, X_i)\}
&= c_n^{\top}\left(A^{-1}-A^{-1} B (B^\top A^{-1} B)^{-1} B^{\top} A^{-1}\right) c_n + o_P(1).
\end{align*}
As a result,
\begin{align*}
2\ell_n(\hat{\theta})-2\ell_n(\theta_0)
&=c_n^\top A^{-1} B (B^\top A^{-1} B)^{-1} B^{\top} A^{-1} c_n + o_P(1).
\end{align*}
By Fact I (with $\theta^*$ replaced by $\theta_0$), routine algebra yields
\[
2\sum_{i=1}^n\log\{1 + \hat{t}_g^\top g_{\hat{\theta}}(Y_i,X_i) + \hat{t}_h^\top \hat{h}^c_i(\tilde{Y}_i, X_i)\}
\rightsquigarrow \mathcal{Z}^\top \mathcal{Z},
\]
where $\mathcal{Z}\sim \cN(0,\Sigma_\mathcal{Z})$ with
\[
\Sigma_\mathcal{Z}=I+ \gamma (J^\top U_\h^{-1}J)^{-1/2}J^\top U_\h^{-1}P g_{\theta_0}\h^{\top}(\cov{\h})^{-1} P \h g_{\theta_0}^{\top} U_\h^{-1}J(J^\top U_\h^{-1}J)^{-1/2}.
\]
Hence, $\mathcal{Z}^\top \mathcal{Z}\rightsquigarrow\sum_{j=1}^p\lambda_j\xi_j^2$, with iid $\xi_j\sim \cN(0,1)$, where $\lambda_j$ are the eigenvalues of $\Sigma_\mathcal{Z}$.
In particular, when $J$ is invertible, $\lambda_j$ are the eigenvalues of 
\begin{align*}
I+ \gamma (J^\top U_\h^{-1}J)^{1/2}J^{-1} P g_{\theta_0}\h^{\top}(\cov{\h})^{-1} P \h g_{\theta_0}^{\top} J^{-\top}(J^\top U_\h^{-1}J)^{1/2}.
\end{align*}

\subsection*{Asymptotic normality of EPI distribution estimators (Theorem~\ref{THM:SUPP:DIST})}

By arguments analogous to those above,
\begin{align*}
\sqrt{n}\{\hat{F}_{\EL}(y,x)-&F(y,x)\}
=\sqrt{n}\{F_n(y,x)-F(y,x)\}\\
&- \begin{pmatrix}
\rho_{gI} \\
\rho_{\h I}
\end{pmatrix}^\top \left(A^{-1}-A^{-1} B (B^\top A^{-1} B)^{-1} B^{\top} A^{-1}\right) c_n + o_P(1).
\end{align*}
where $F_n(y,x)=n^{-1}\sum_{i=1}^n I(Y_i\le y, X_i\le x)$, $\rho_{gI}=E\big(g_{\theta^*}(Y,X)I\{Y\le y, X\le x\}\big)$, and $\rho_{\h I}=\cov\big(\h(\tY,X),I\{Y\le y, X\le x\}\big)$.

The proof of Fact I implies that
\begin{align*}
c_n=\begin{pmatrix}
\sum_{i=1}^n n^{-1/2} g_{\theta^*}(Y_i,X_i) \\
\sum_{i=1}^n \sqrt{n}(n^{-1}-N^{-1})\{\h(\tY_i,X_i)-P\h\}+\sum_{i=n+1}^{n+m}(-\sqrt{n}N^{-1})\{\h(\tY_i,X_i)-P\h\}
\end{pmatrix}+o_P(1).
\end{align*}
In addition,
\[
\sqrt{n}\{F_n(y,x)-F(y,x)\}=\sum_{i=1}^n n^{-1/2}\{I(Y_i\le y, X_i\le x)-F(y,x)\}.
\]
It is straightforward to show that
\begin{align*}
\begin{pmatrix}
c_n\\
\sqrt{n}\{F_n(y,x)-F(y,x)\}
\end{pmatrix}\rightsquigarrow N\Big(0,\Xi\Big),
\end{align*}
where
\[
\Xi=\begin{pmatrix}
P g_{\theta^*} g_{\theta^*}^\top & (1-\gamma) P g_{\theta^*}\h^{\top} & \rho_{gI}\\
(1-\gamma)P \h g_{\theta^*}^{\top} & (1-\gamma)\cov{\h} & (1-\gamma)\rho_{\h I}\\
\rho_{gI}^\top & (1-\gamma)\rho_{\h I}^\top & F(y,x)\{1-F(y,x)\}
\end{pmatrix}.
\]
Routine algebra yields $\sqrt{n}\{\hat{F}_{\EL}(y,x)-F(y,x)\}\rightsquigarrow \cN(0,\tau^2)$, where, with $\Pi_{U;\h}=U_\h^{-1}-U_\h^{-1}J(J^\top U_\h^{-1}J)^{-1}J^\top U_\h^{-1}$, $a_\h=\rho_{gI}-P g_{\theta^*}\h^{\top}(\cov{\h})^{-1}\rho_{\h I}$ and $b_\h=P g_{\theta^*}\h^{\top}\Pi_{U;\h} a_\h-\rho_{\h I}$, we have
\begin{align*}
\tau^2=F(y,x)\{1-F(y,x)\}-\rho_{\h I}^\top (\cov{\h})^{-1}\rho_{\h I}-a_\h^\top\Pi_{U;\h} a_\h+\gamma b_\h^\top(\cov{\h})^{-1}b_\h.
\end{align*}
In particular, \(\tau^2 \leq F(y,x)\{1-F(y,x)\}\) if \(\gamma b_\h^\top(\cov{\h})^{-1}b_\h\leq \rho_{\h I(y, x)}^\top (\cov{\h})^{-1}\rho_{\h I(y, x)}+a_\h^\top\Pi_{U;h} a_\h\). When $J$ is invertible (so that $\Pi_{U;\h}=0$), this simplifies to
\[
\tau^2=F(y,x)\{1-F(y,x)\}-(1-\gamma)\rho_{\h I}^\top (\cov{\h})^{-1}\rho_{\h I}.
\]

\putbib
\end{bibunit}

\end{document}